\documentclass[apj,onecolumn,12pt]{emulateapj}
\usepackage{amsmath}
\input psfig.tex

\begin{document}

\shorttitle{VIRUS Optical Fibers}

\title{The Influence of Motion and Stress on Optical Fibers}

\author{Jeremy D. Murphy\altaffilmark{a,c}, Gary J. Hill\altaffilmark{b},
Phillip J. MacQueen\altaffilmark{b},\\
Trey Taylor\altaffilmark{b}, Ian Soukup\altaffilmark{e},
Walter Moreira\altaffilmark{b}, Mark E. Cornell\altaffilmark{b}, John Good\altaffilmark{b},\\
Seth Anderson\altaffilmark{c}, Lindsay Fuller\altaffilmark{c},
Hanshin Lee\altaffilmark{b}, Andreas Kelz\altaffilmark{d}, Marc Rafal\altaffilmark{b},\\
Tom Rafferty\altaffilmark{b}, Sarah Tuttle\altaffilmark{b}, Brian Vattiat\altaffilmark{b}}

\altaffiltext{a}{Princeton University, Department of Astrophysical Sciences,
4 Ivy Lane, Peyton Hall, Princeton, NJ 08544}
\altaffiltext{b}{The University of Texas, McDonald Observatory,
2515 Speedway C1402, Austin, Texas, USA 78712-1206}
\altaffiltext{c}{The University of Texas, Department of Astronomy,
2515 Speedway C1400, Austin, Texas, USA 78712-1205}
\altaffiltext{d}{Leibniz-Institut f$\ddot{\rm u}$r Astrophysik,
An der Sternwarte 16, 14482 Potsdam, Germany}
\altaffiltext{e}{The Center for Electromechanics, The University of Texas,
1 University Station PRC R7000, Austin, Texas 78712}

\email{jdm@astro.princeton.edu}

\begin{abstract}

We report on extensive testing carried out on the optical fibers for the VIRUS instrument. The primary result of this work explores how 10+ years of simulated wear on a VIRUS fiber bundle affects both
transmission and focal ratio degradation (FRD) of the optical fibers. During the accelerated lifetime tests we continuously monitored the fibers for signs of FRD. We find that transient FRD events were common during the portions of the tests when motion was at telescope slew rates, but dropped to negligible levels during rates of motion typical for science observation. Tests of fiber transmission and FRD conducted both before and after the lifetime tests reveal that while transmission values do not change over the 10+ years of simulated wear, \emph{a clear increase in FRD is seen in all 18 fibers tested.} This increase in FRD is likely due to microfractures that develop over time from repeated flexure of the fiber bundle, and stands in contrast to the transient FRD events that stem from localized stress and subsequent modal diffusion of light within the fibers. There was no measurable wavelength dependence on the increase in FRD over 350~nm to 600~nm. We also report on bend radius tests conducted on individual fibers and find the 266 $\mu$m VIRUS fibers to be immune to bending-induced FRD at bend radii of R~$\ge 10$~cm. Below this bend radius FRD increases slightly with decreasing radius. Lastly, we give details of a degradation seen in the fiber bundle currently deployed on the Mitchell Spectrograph (formally VIRUS-P) at McDonald Observatory. The degradation is shown to be caused by a localized shear in a select number of optical fibers that leads to an explosive form of FRD. In a few fibers, the overall transmission loss through the instrument can exceed 80\%. These results are important for the VIRUS instrument, and for both current and proposed instruments that make use of optical fibers, particularly when the fibers are in continual motion during an observation, or experience repeated mechanical stress during their deployment.

\end{abstract}

\section{INTRODUCTION}
\label{intro}

First proposed for use in astronomical instrumentation by \citet{ang77} optical fibers have revolutionized the field over the past 3 decades. This revolution has come about due to the flexibility optical fibers offer in re-routing the light from thetelescope focal plane to a more convenient location. The gain for this technological advance is clear as issues of instrument weight, size, stability, and temperature control are made largely obsolete. However, this advantage comes at a cost as fibers constitute an added element in the instrument's optical path.  Moreover, optical fibers are not entirely stable light guides, and while their characteristics have been studied in a comprehensive and systematic way by several groups \citep{ram88, sch03, cra08, pop10a}, little work has been focused on their behavior during periods of motion and accumulated stress, with notable and valuable exceptions \citep{cra88, cla89, avi98, bry10, hay11, bry11}. As both current \citep{ber04, kel04, smi04, rot05, kel06b, tut08, wil10} and proposed instrumentation \citep{nav10, sau10} is making heavy use of optical fibers, with many instruments requiring repeated motion of the fiber optics, a clear understanding of their properties under these conditions is needed.

Transmission and focal ratio degradation (FRD) are the two properties
of optical fibers that are generally of most interest for astronomical
instrumentation. Unlike transmission values, which are now routinely
supplied by the fiber vendor and, over a wide wavelength range, are
generally repeatable in the lab, FRD measurements are typically left
to the individual groups responsible for the instrument. This is in
large part due to two related issues in quantifying FRD. One, due to
the low levels of light involved in measuring FRD, and the significant
potential for systematic errors introduced by the experimental set-up,
accurate measurements of FRD are challenging. Two, there is not one
single source of FRD, but rather several affects of fiber polishing,
mounting, and on-telescope application that drive the various causes of
a divergent output light cone that we generally refer to under the
generic term of FRD.\citep{hay11} However, despite these challenges, several
groups have quantified FRD via a number of different techniques.
\citep{ram88,cra88,sch98,avi98,car94,sch03,cra08,mur08,hay08,bru10,pop10b} 

FRD is typically defined as any increase in the
output angle (i.e output f-ratio, hereafter f/out) of
light when compared to the input angle. In an ideal fiber, the
input f/ratio (f/in) will be preserved through the fiber, with no rays
being scattered to larger output angles. In this case, f/in =
f/out and the fiber does not suffer from FRD. For this work we
broaden the definition of FRD to include the scattering of rays to
lower angles as well as higher ones. Another way to express this modified
definition of FRD is by saying that no FRD corresponds to no radial
scattering of the light injected into a fiber. We employ
this modified definition of FRD because VIRUS uses a Schmidt camera
design \citep{hil10}. With such a camera design, light thrown into 
the central obscuration (see Figure \ref{spots}) can land on
the back of the CCD, leading to both loss of overall transmission and
the possibility of scattered light. Note that while there are
multiple sources of FRD seen in our various tests, we leave the
majority of the discussion of the sources of FRD for \S \ref{discuss},
and focus on its quantifiable effects. Therefore we will use the
generic term ``FRD'' throughout this work to mean any affect that
leads to radial scattering of light within a fiber.

In this work we report on tests performed on both individual fibers
and fiber bundles for use in the Visible Integral-field Replicable
Unit Spectrograph (VIRUS) \citep{hil10}. VIRUS is a fiber-fed
spectrograph currently under construction to carry out the Hobby
Eberly Telescope Dark Energy eXperiment (HETDEX)
\citep{hil08a}. Composed of 75 integral field units (IFU), each 
$\sim 22$~m long and made of 448 optical fibers, \citep{kel06} VIRUS
will employ the use of fiber optics in astronomical instrumentation on a
unprecedented scale. As the Hobby-Eberly Telescope (HET) tracker is in
continual motion
during a science observation, a clear understanding of both the
behavior of fibers while in motion, and the effect several years of
motion have on the fiber properties, is essential.

The outline of this paper is as follows. In \S \ref{methods} we
provide an overview of our experimental set-ups and method of analysis
of FRD. Bend radius tests conducted on 3 different fiber types are
detailed in \S \ref{bendtests}. Our main results are found in \S
\ref{lifetime}, where we describe the lifetime fiber tests carried out
on a single VIRUS fiber bundle, and the changes we observed over the
course of the tests. Then, in \S \ref{mspec}, we report on a
specific case for the current science IFU (often referred to as a
``fiber bundle'' throughout this work) in use on the Mitchell
Spectrograph (formally VIRUS-P) \citep{hil08b} where FRD in 18 fibers
has increased over time, in some cases substantially. In certain
fibers the increase in FRD is extreme, leading to a $\ge 80$\%\ drop
in overall fiber transmission. In \S \ref{discuss} we discuss the
implications each of these results has for the VIRUS instrument and other
projects which employ fibers, particularly if the fibers are mobile
or experience repeated motions or localized stress during their deployment.

\section{TESTING METHODS}\label{methods}

\begin{figure}
\begin{center}
\begin{tabular}{c}
\includegraphics[scale=0.72,trim=50 70 50 0,clip,angle=0]{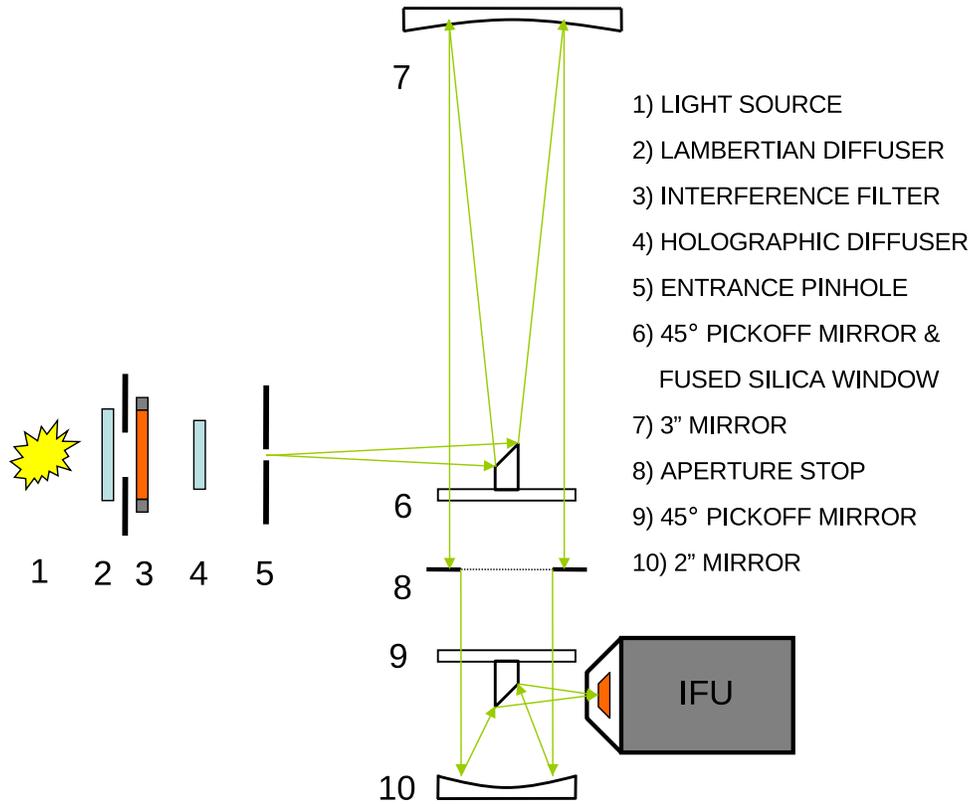}
\end{tabular}
\end{center}
\caption[Schematic of the reconfigured fiber test bench.]{A schematic
  of the fiber optic test bench used for the transmission and FRD
  measurements presented here. The primary optical elements are the
  same as in Murphy et al. (2008; hereafter M08)\cite{mur08}, with
  slight changes in the illumination system. The arrows indicate the
  path of light through the test bench. The light source (\#1) is
  diffused down to achieve a very even illumination on the 700 $\mu$m
  pinhole (\#5) which sits at the focus of the first mirror (\#7). The 
  first of two $45^\circ$ pickoff mirrors (\#6) send the light from the
  pinhole to the first mirror where it comes back collimated. An iris
  (\#8) sets the input f-ratio, and a second mirror (\#10) and
  $45^\circ$ pickoff mirror (\#9) focuses the light down onto the end of an
  individual fiber. The final spot size we couple into the fiber is
  $\sim 170$~$\mu$m, which slightly underfills the 266~$\mu$m diameter
  fibers and is similar to a 1.0 arcsec point source for the VIRUS
  instrument. 
  \label{bench}}
\end{figure}

\begin{figure}
\begin{center}
\begin{tabular}{c}
\includegraphics[scale=0.20,trim=0 0 0 0,clip]{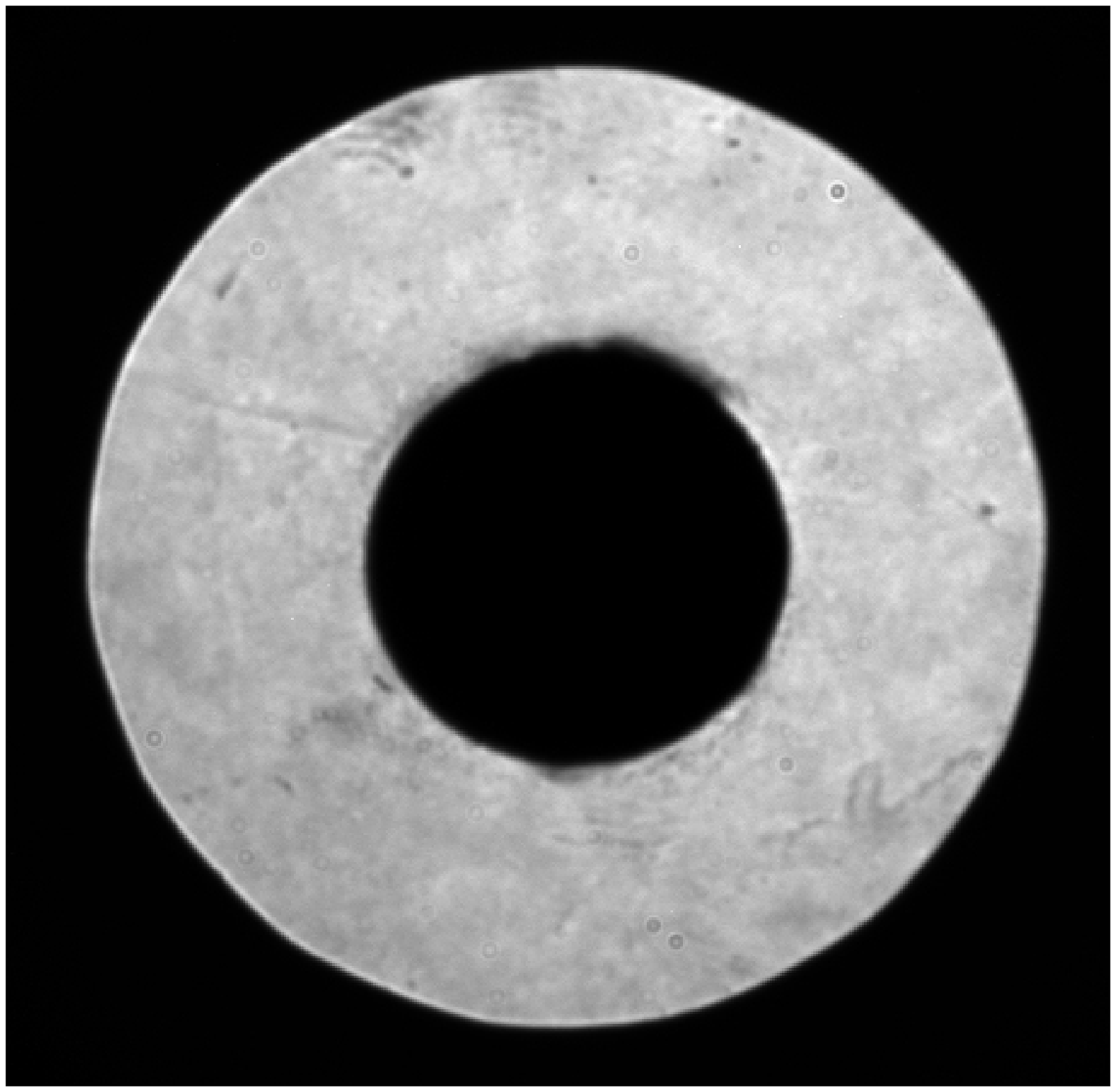}
\includegraphics[scale=0.286,trim=0 23 0 0,clip]{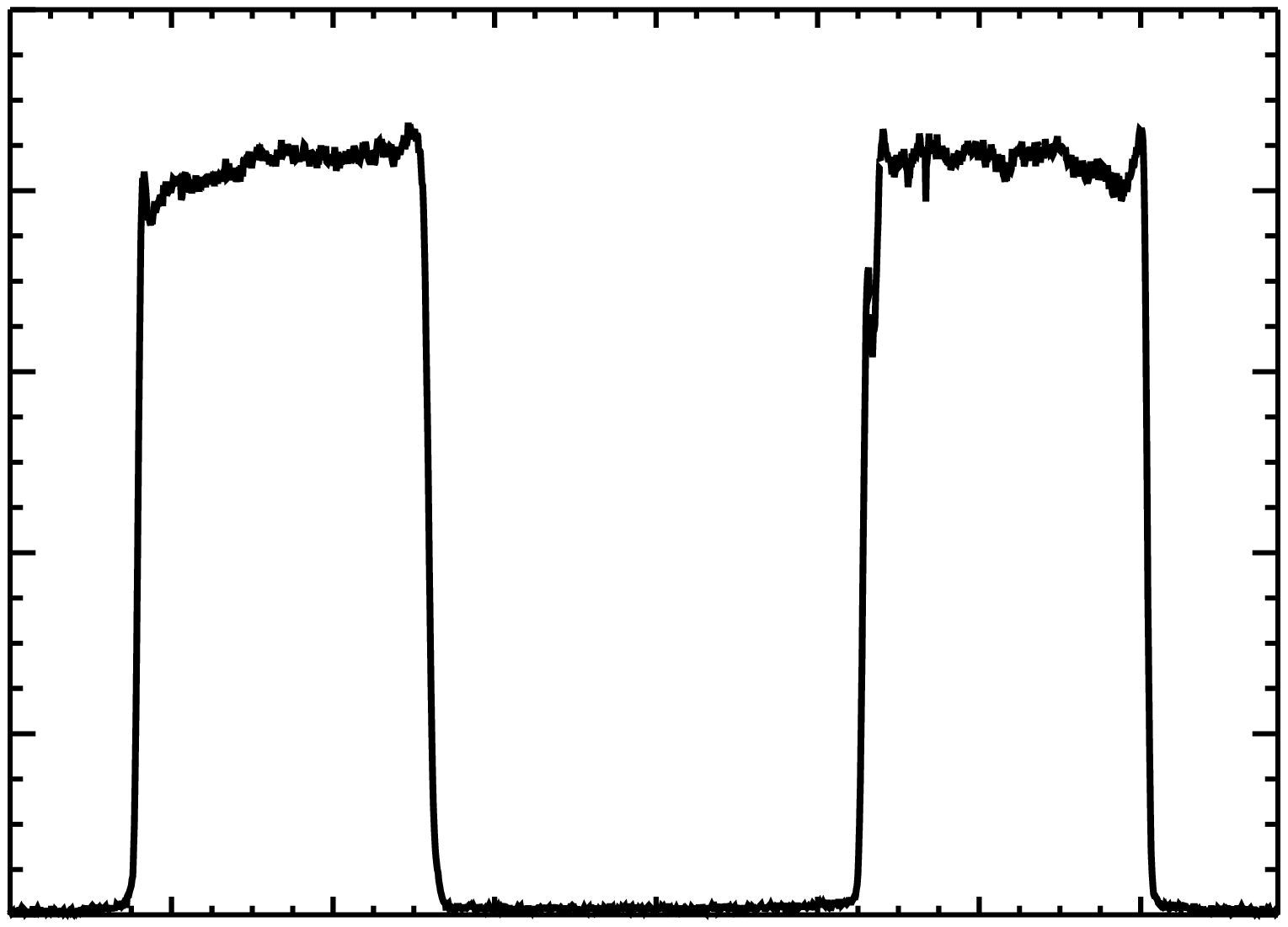}
\includegraphics[scale=0.375,trim=0 0 0 0,clip]{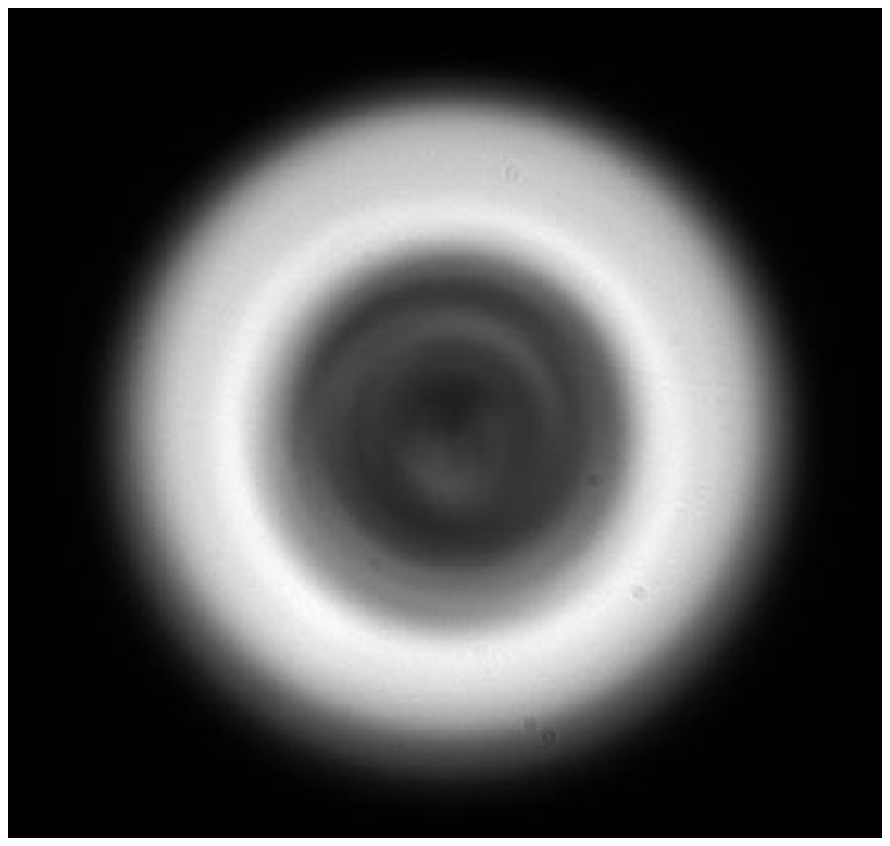}
\includegraphics[scale=0.287,trim=0 23 0 0,clip]{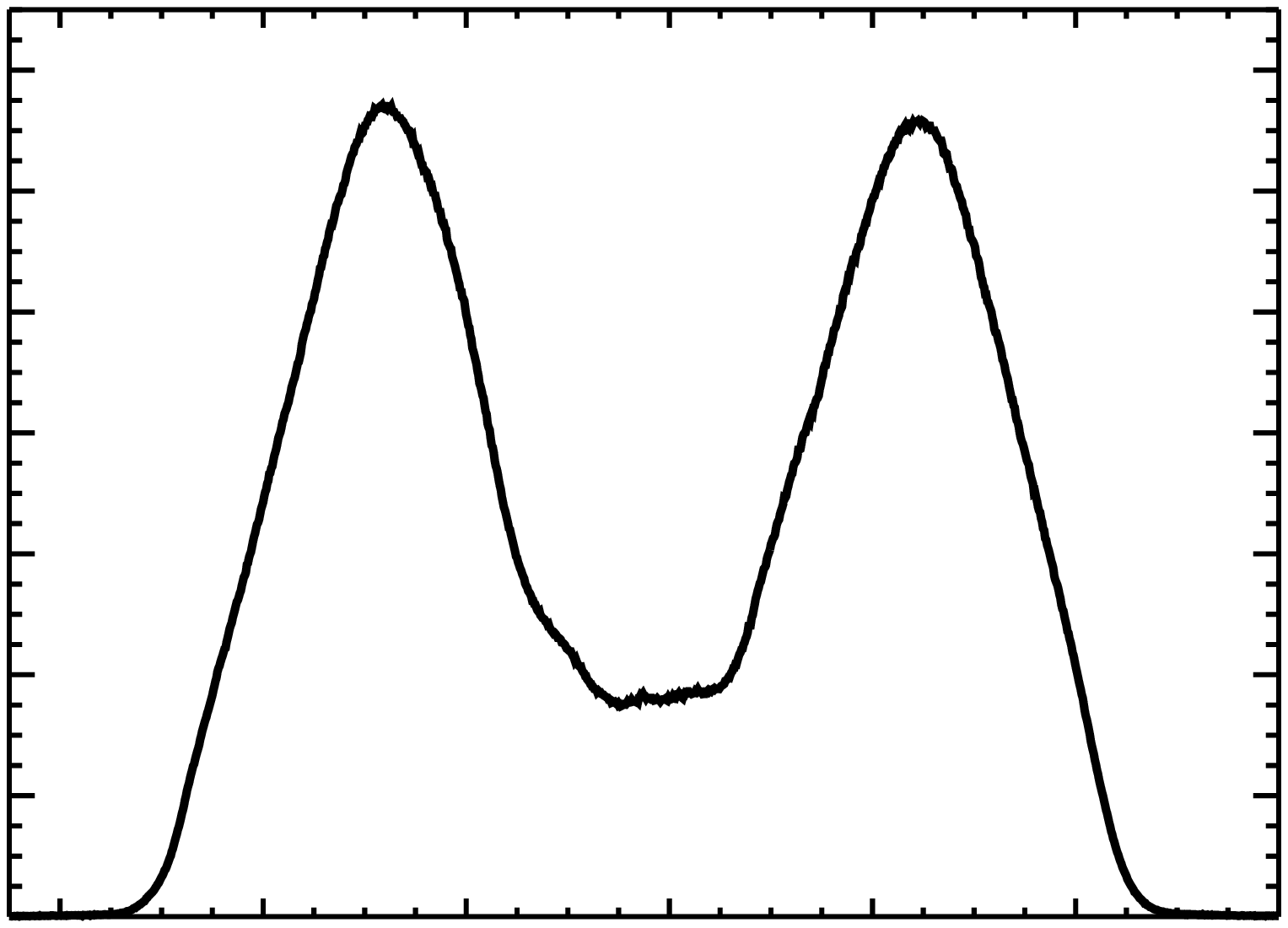}
\end{tabular}
\end{center}
\caption[Far-field images and cross-sectional profiles of the input
  and output of the fiber optic test bench.]{Far-field image and
  cross-sectional profile of the input spot (left 2 figures) coming
  from our fiber optic test bench. Shown in the right 2 images are
  similar frames, but showing a typical output spot \emph{after} the
  light has passed through an optical fiber. The radial scattering of
  light to both higher and lower output angles, which we define as FRD
  for this work, is clearly evident. An encircled energy (EE)
  analysis of the far-field image at each of 7 discrete camera
  positions returns an output focal ratio (f/out) measurement
  at 1\%\ increments between an EE of 5\%\ and 95\%. These values are
  then used to reconstruct the output light cone and quantify the FRD
  of a single fiber.
  \label{spots}}
\end{figure}

Our primary tests of transmission and FRD follow the method laid out
in \citet{mur08} (hereafter M08). We provide an
overview of our test set-up and method for measuring transmission and
FRD here, leaving the interested reader to reference M08 for further
details. Since the work presented in M08 was conducted, the test bench
has been automated. During this modification process we improved the
diffusion of the input light source and removed the one lens
(element \#5 in Figure 3 of M08) as other improvements to the test
bench made this focusing lens unnecessary. The layout of the rest of
the test bench remains unchanged from M08, and is shown in Figure
\ref{bench}. Two 45-degree pickoff mirrors (elements \#6 and \#9) 
create a shadow at $\sim$f/8 that mimics the secondary of a typical
telescope. In Figure \ref{spots} we show far-field images and
cross-sectional profiles of the spot that the test bench creates at
its focus (left 2 images) and the output after the light has passed
the length of a fiber (right 2 images). The central 
obscuration created by the pickoff mirrors has proven incredibly
informative in our analysis of FRD as it allows us to explore FRD
effects that radially scatter light into the central obscuration. An
iris placed in the collimated light path (element \#8) is used to set
the f/in of the test bench. A cone-style FRD analysis and transmission
measurement are conducted at 9 discrete wavelengths ($\sim 10$~nm
FWHM) between 350 nm and 600 nm. This method involves using the second
mirror (element \#10) to focus the collimated light from the test
bench onto the end of an individual fiber. The fiber is placed at the
focus of the second mirror and therefore sees a cone of light input at
f/3.65. This light is very similar to the light seen by the fibers on
the telescope.

\begin{figure}
\begin{center}
\begin{tabular}{c}
\includegraphics[scale=0.70,trim=0 0 0 0,clip]{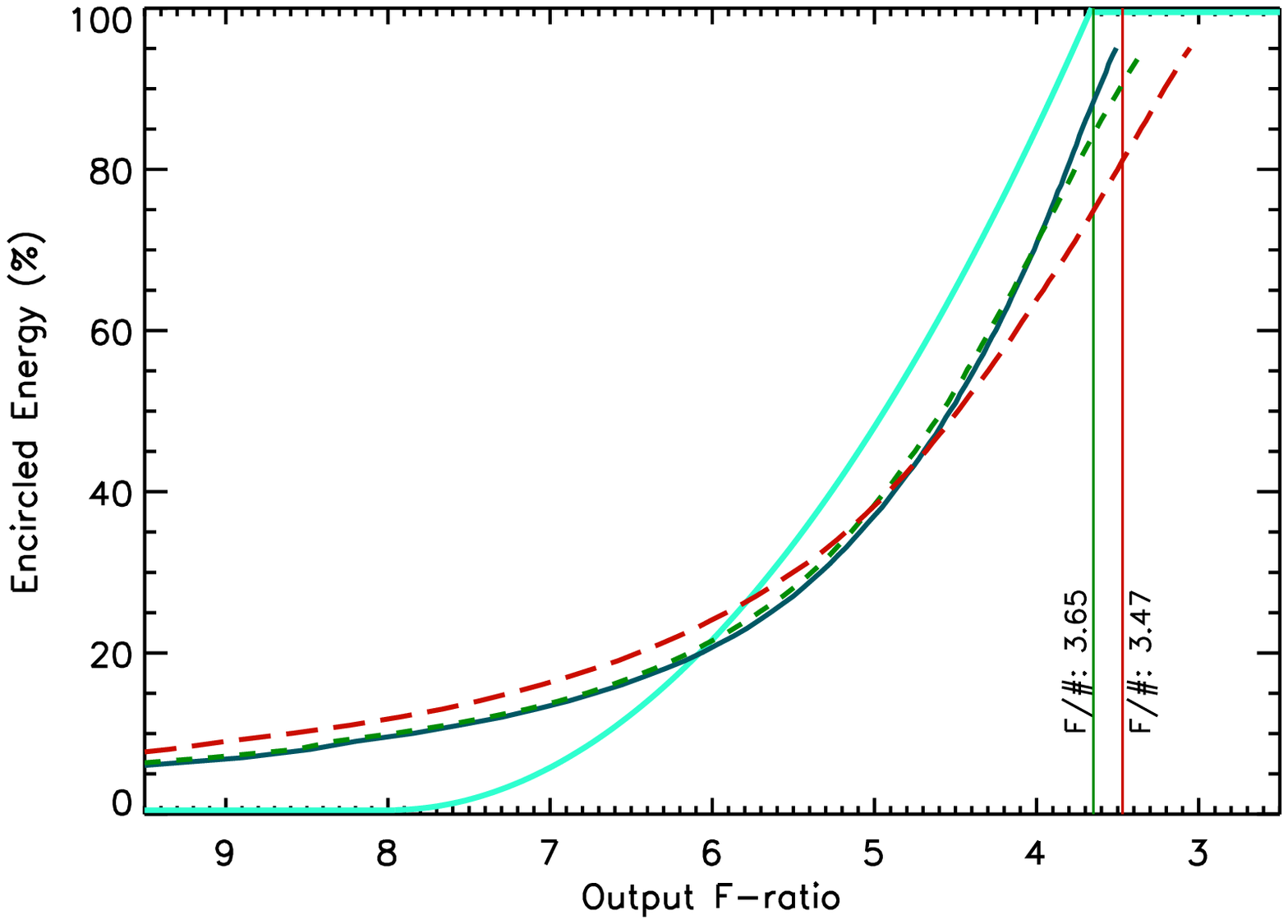}
\end{tabular}
\end{center}
\caption[Encircled energy vs. f/out for a set of typical optical
  fibers.]{Encircled Energy (EE) as a function of f/out 
  for three fibers from the lifetime test fiber bundle showing a range
  of FRD. By plotting a wide range of EE vs f/out (rather than quoting
  the f/out value at a single EE value) we are able to map out the effects
  of FRD over the entire output spot. The vertical green line marks our
  f/in value of f/3.65 while a second red line denotes the
  f/3.47 (95\% EE) tolerance for the VIRUS spectrograph. The light teal
  line plots the theoretical EE vs. f/out in the case of no
  FRD; EE remains flat at 0\%\ until the inner edge of the central
  obscuration is hit, then climbs as R$^2$, in accord with the
  increase in area of the integrated circle that defines our EE, then
  flattens out again once at 100\%\ when the outer edge of the spot is
  reached. The solid blue line plots data from a fiber exhibiting
  typical levels of FRD. The short-dashed
  green line shows more FRD, yet with the predominant increase
  coming as faster rays (i.e. light scattered outward). For the fiber
  plotted as a long-dash red line we see FRD scattering light into
  both faster and slower rays (i.e. outward \emph{and} into the
  central obscuration). A 
  critical point to note is the way in which the fiber exhibiting the
  worst FRD \emph{crosses over} the other fibers. This cross-over
  is characteristic of an increase in FRD, as light is thrown
  into both the central obscuration and outer halo. Two of the three
  fibers plotted do not meet the specification of 95\% EE within
  f/3.47. However, the lifetime test fiber bundle was not polished to
  specification and neither input or output cover-plate and
  index-matching gel was installed for these tests. 
  \label{frdvee}}
\end{figure}

The first two frames in Figure \ref{spots} show a typical far-field
image and cross-sectional profile of the ``input'' spot (i.e. the cone
of light the test bench creates and that is coupled into an optical
fiber). The input spot is used to both confirm the input f-ratio of
the fiber test bench and provide a ``baseline'' flux for the transmission
measurements. The light source is very stable ($\sim 0.05$\% variation
over several hours) and therefore provides the reference point for our
transmission measurements. The right two frames in Figure \ref{spots}
show a far-field image and cross-sectional profile for the ``output''
spot (i.e. the light \emph{after} it has passed the length of an
optical fiber). The central obscuration, simulating the secondary
mirror of a typical telescope, is clear in all the output images, with
some amount of FRD throwing light both into the central obscuration
and the outer halo. By stepping the camera over 7 small (4 mm), well
controlled increments along the optical axis of the fiber, and
capturing images of the far-field at each camera step, we can map out
the output light cone. At each camera position we take a series of
exposures and darks for each of the 9 interference filters. These
frames, once referenced to the corresponding baseline frame, provide 7
measurements of the transmission (one from each camera position) at
each of our 9 bandpasses.

In order to quantify the effect of FRD over the entire fiber
profile we consider the f/out at a wide range of EE, rather than quoting
a single value (e.g. 95\% EE). The measure of FRD for a single fiber is
accomplished by determining the radius, in pixels, for a range of
encircled energy (EE) values from 5\% to 95\%, in 1\% increments, for
each of the 7 camera positions and 9 wavelengths. A line is fit
through the 7 radius values at each EE value. This set of lines then
defines the output angle of the light from a single fiber for the 90
different EE values we've measured. By knowing the camera step size
and physical CCD pixel size we can determine the f/out, in absolute
terms, and thus the FRD of the fiber. Figure \ref{frdvee} shows an
example of this analysis for three fibers in the fiber bundle used for
the lifetime tests (see \S \ref{lifetime}), and plots EE against f/out
for three fibers exhibiting various degrees of FRD. The input f-ratio
(f/3.65) is shown as a vertical line. The light teal line represents
what a fiber experiencing no FRD would look like. The solid blue line
shows a fiber with good FRD characteristics; there is a low level of
light thrown into the center (as exhibited by the lower values of EE
at higher f/out), and less light thrown outwards (as seen in the
steepness of the increase in EE at lower f/out). The short-dash green
fiber shows a slight worsening of FRD, primarily in the outer
halo. The long-dash red fiber shows significant FRD, with elevated
levels of light both in the central obscuration and the outer
halo. Figure \ref{frdvee} reveals an interesting effect of a
well-studied phenomena of optical fibers. Rays incident on a fiber at
higher f/in suffer from higher FRD (e.g. \citet{ram88,pop10a}). As the rays near the central obscuration are  
incident at higher f/in ($\sim$~f/8) than those near the outer edge of
the input spot ($\sim$~f/3.6), they suffer from more FRD than those
near the edge of the input spot. Therefore, the severity of the FRD
experienced by a ray is a function of it's incident angle, which
changes smoothly from the center of the input spot to the outer
edges. If the strength of the FRD on the central rays is not strong 
enough to send the light entirely outside f/3.65, you should expect to
see a piling-up effect of the light in the output spot. We see such an
effect, as evidenced by the blue FRD vs. EE curve being steeper
between $\sim$f/5 to $\sim$f/3.5 than the curve showing the
theoretical light profile of a fiber experiencing no FRD. You can also
see this effect directly, although not in a quantified way, by
visually inspecting the far-field images of the output spots (Figure
\ref{spots}).

\section{BEND RADIUS TESTS}\label{bendtests}

\begin{figure}
\begin{center}
\begin{tabular}{c}
\includegraphics[scale=0.45,trim=0 0 0 0,clip]{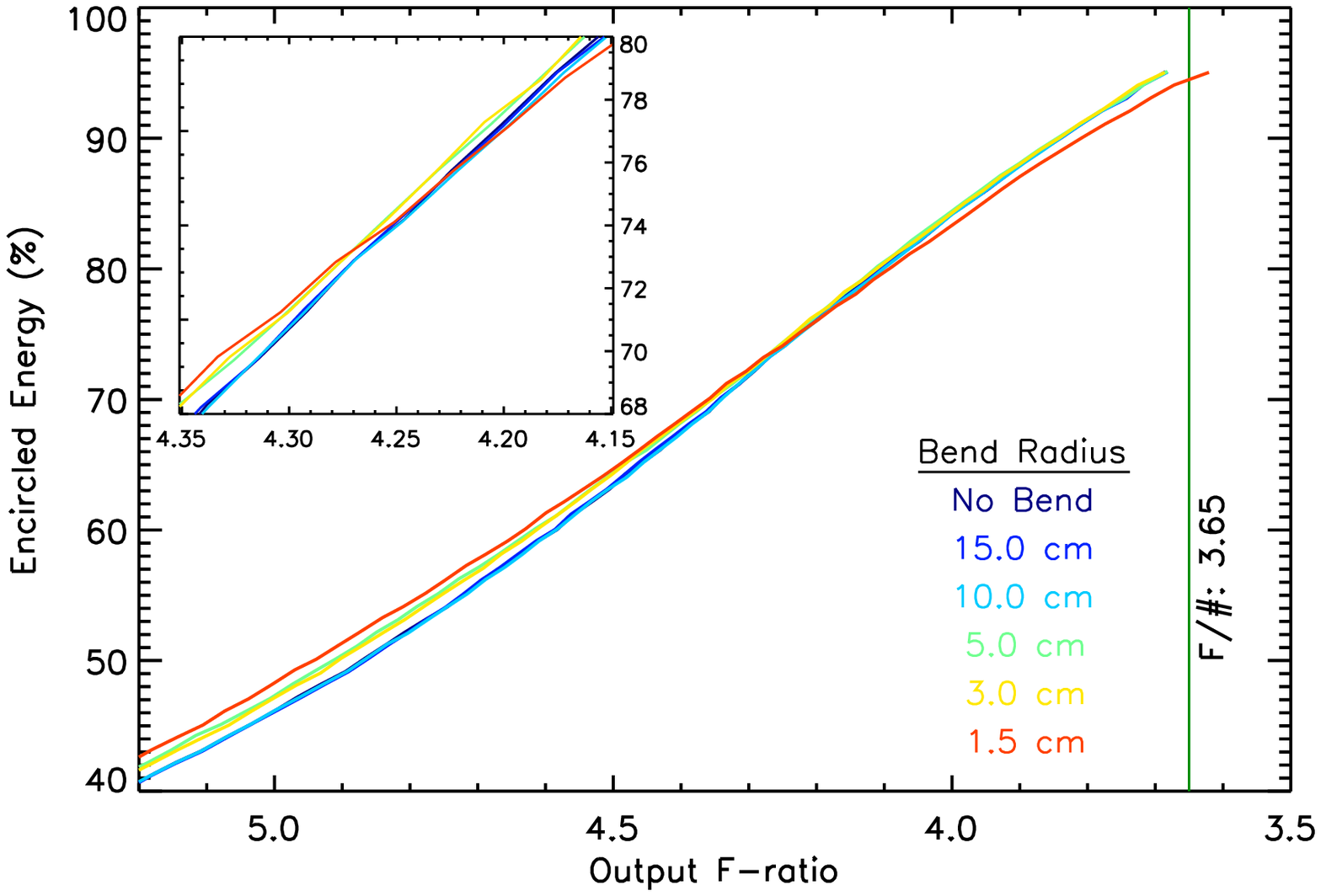}
\includegraphics[scale=0.45,trim=0 0 0 0,clip]{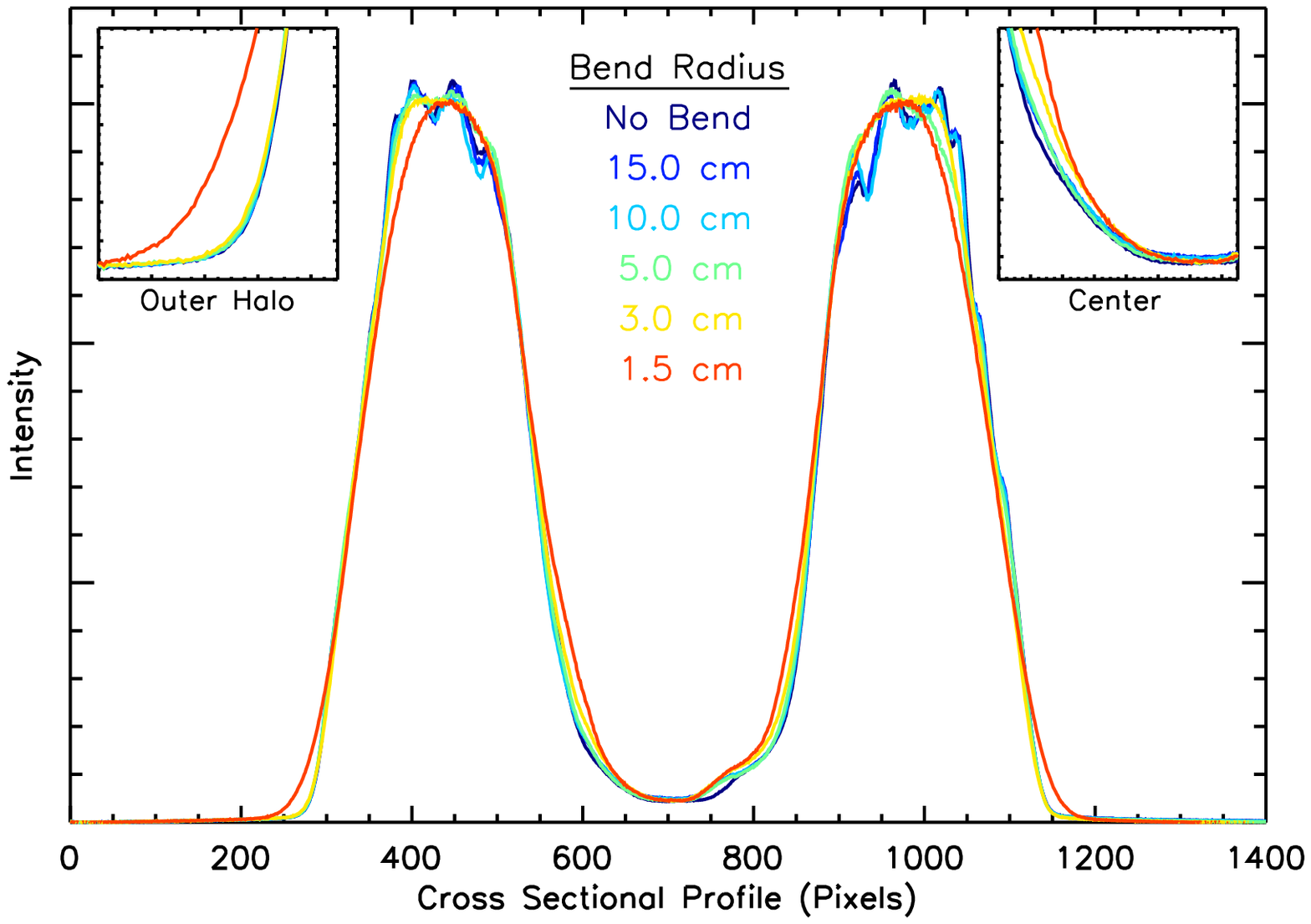}
\end{tabular}
\end{center}
\caption[Bend radii tests showing fiber FRD is not present for bend
  radii greater than 10 cm.]{Left: EE vs f/out for 6 bend radii tested
  on a single fiber (Fibertech AS266/292UVPI/318). The tests were
  performed on single $\sim 6$~m fibers, mounted in standard ferrules
  at both ends, and polished by hand. A single loop was placed at the
  midpoint of the fiber. No measurable increase in FRD occurs between
  the no bend, 15~cm, and 10~cm bend radii. At the bend radii of 5~cm
  and 3~cm a small increase in FRD is seen, \emph{yet only into the
  central obscuration.} We believe this light comes from the slower
  ($\sim$f/8) rays near the central obscuration which are more prone
  to being scattered by stress-induced modal diffusion. At
  R$_{\mathrm{bend}}$~=~1.5~cm a further increase in FRD is observed,
  this time with scattering observed into \emph{both the 
  central obscuration and the outer halo}. The inset in the
  upper-right shows a close-up of the cross-over of the EE vs. f/out
  lines characteristic of FRD. Right: Cross-sectional
  profiles of the far-field images for all 6 bend radii. Note that not
  only does bending-induced FRD scatter rays into the outer halo and
  central obscuration, but generally smooths the cross-sectional
  profile. The two insets show a closeup of the outer halo and
  central obscuration.
  \label{bend}}
\end{figure}

In order to set a fiber bend radius limit for the VIRUS instrument we
conducted a series of tests to explore the influence of fiber bending
on FRD. Three different fiber types were tested: CeramOptec UV265/292P/320,
Fibertech AS266/292UVPI/318 and F~\&~T 266/292. The three, 10 m length
fibers were inserted into furcation tubing, mounted with standard
SMA905 fiber connectors, then both ends were polished by hand and
tested for their bend radius characteristics. First, each fiber was
tested twice for FRD without any bends in the fiber, once through each
end of the fiber. Then, the fiber was laid flat on a table, with a
single loop placed into the center of the fiber. Care was taken not to
add any extra twists into the fiber when making the loop. The fiber
was then tested for FRD at increasing bend radii from R = 15 cm to R =
1.5 cm. The fiber was then retested in the same manner, yet coupling
light into the other end of the fiber.

As the results for all three fibers are very similar, we have plotted
the results from just the Fibertech fiber in Figure \ref{bend}. Note
the high quality of the end polish achieved for these tests as
evidenced by the very low level of light in the central obscuration
(Figure \ref{bend}, right). The effect of FRD with bend radius is
subtle, yet clearly evident. For the first three bend radii (none,
15~cm and 10~cm) there is no measurable increase in FRD or change in
the cross-sectional profile of the far-field output spot. Then, at
both 5~cm and 3~cm bend radii, we see the first evidence of FRD, and
note an interesting affect; \emph{FRD is evident in the central
  obscuration, but not in the outer halo.} This stems from the same
affect described in the previous section, with the $\sim$f/8 rays near
the central obscuration suffering from stronger FRD. It is only when
the bend radius reaches 1.5~cm that we see clear FRD into the outer
halo. It is also worth noting the mode mixing that occurs with bend
radius, as seen in the smoothing of the cross-sectional profiles,
plotted to the right in Figure \ref{bend}. 

\section{LIFETIME FIBER TESTS}\label{lifetime}

\begin{figure}
\begin{center}
\begin{tabular}{c}
\includegraphics[scale=0.336,trim=0 0 0 0,clip]{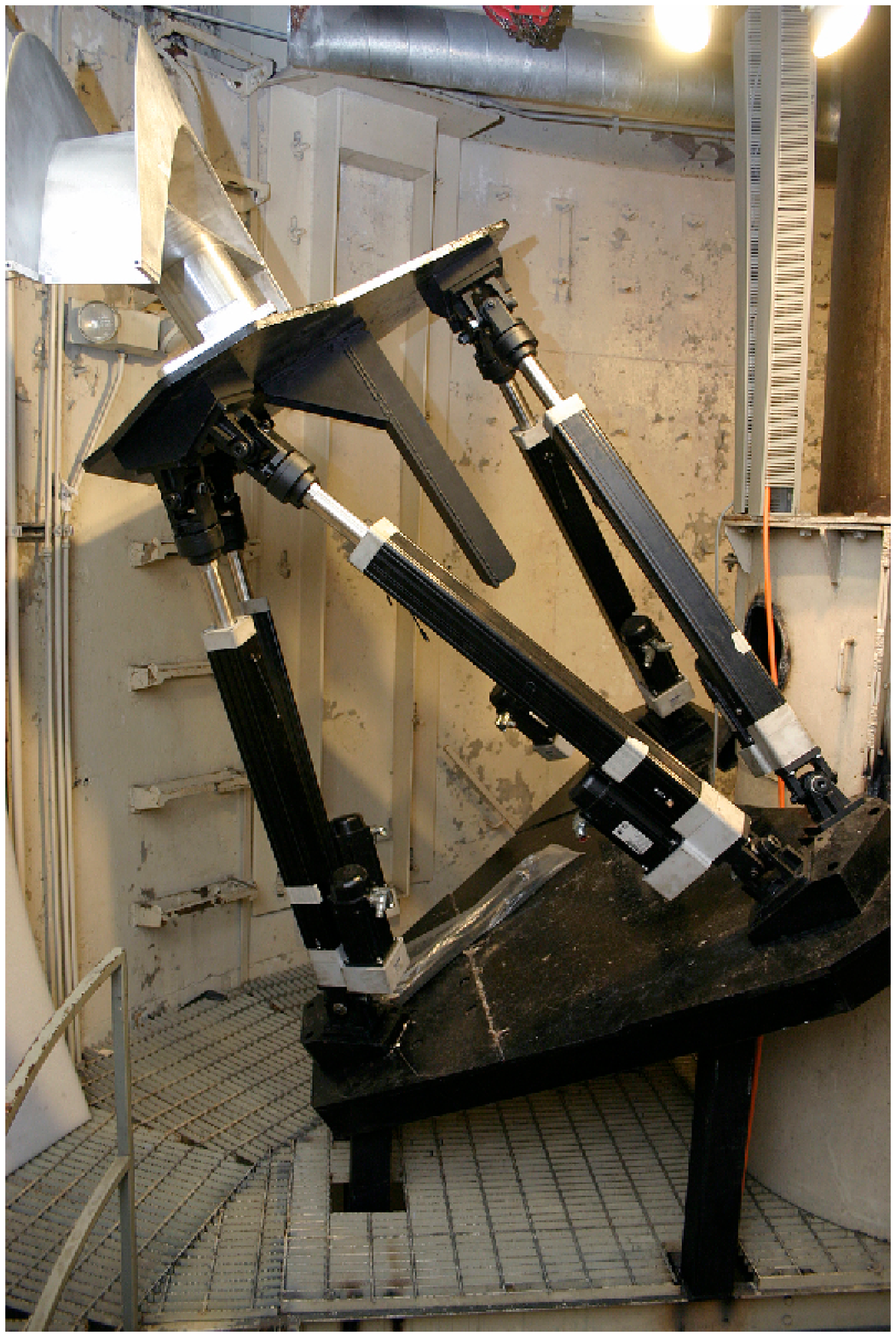}
\includegraphics[scale=0.0675,trim=0 0 0 0,clip]{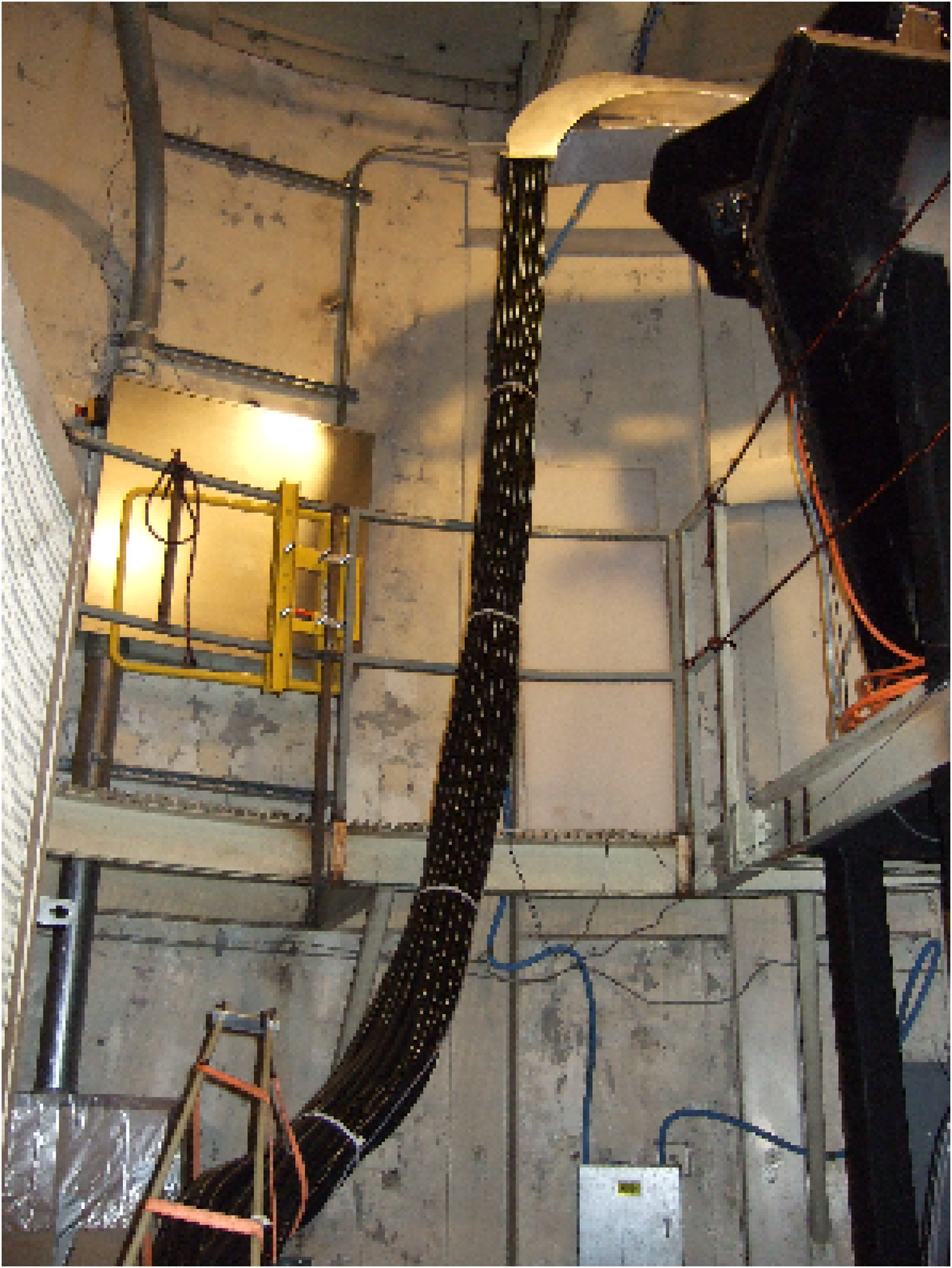}
\includegraphics[scale=0.335,angle=0,trim=0 0 0 0,clip]{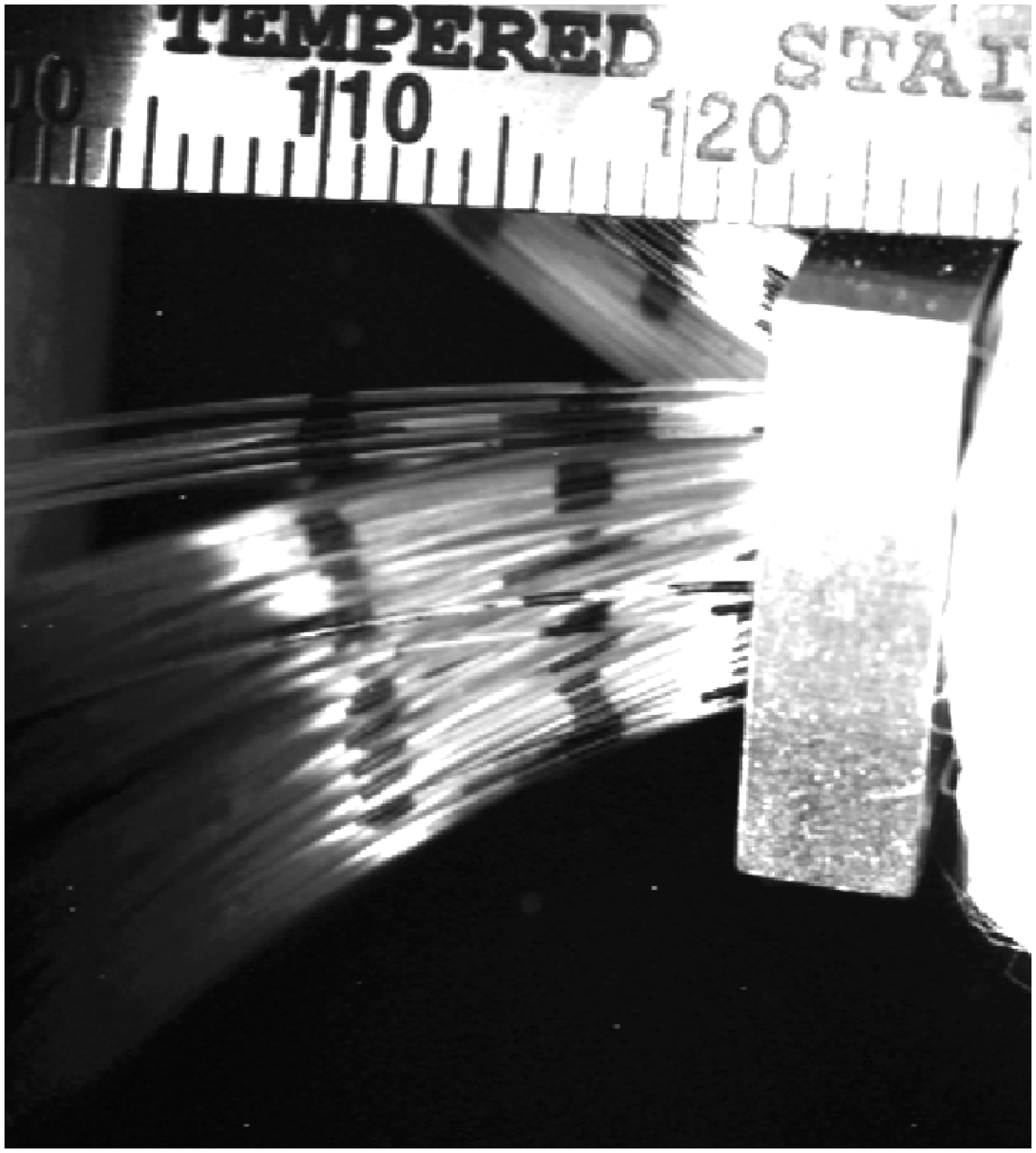}
\end{tabular}
\end{center}
\caption[Images of the lifetime test set-up.]{Left: An image of the
  test platform used in the lifetime tests to simulate the translation
  and rotation of the Prime Focus Instrument Package (PFIP), which
  constitutes the secondary of the HET (see Figure 1 in \citet{hil10}). The fibers and input light 
  source, which gets mounted to the black, L-shaped support below the
  platform, had not been installed when this image was taken. Center:
  An image taken from below the test platform, looking up. The black
  dummy fiber bundles, used to simulate the pressure and collective
  action of all the VIRUS fiber bundles on the test bundle, have
  been installed and are supported at the first strain relief (top-right in 
  image). The test fiber bundle runs through the center of the dummy 
  fiber bundles and is therefore not visible in this image. Right: An
  image taken at the output end of the fiber bundle, as the fibers
  break out of the fiber conduit and before they are arrayed into
  the output slits (see the central image of Figure 6 in Hill, \cite{hil10}). The motion of the fibers relative to their
  protective conduit was monitored in this fashion for the duration of
  the lifetime tests. 
  \label{setup}}
\end{figure}

There are two critical aspects of the VIRUS IFU performance we want to
understand and that were not explored in M08 or in other published
work. First, how do optical fibers behave while in motion, and do these
effects depend on the rate of fiber motion, if at all? This
understanding is necessary for VIRUS as the HET tracks its targets via
motion of a suite of mirrors at prime focus (called the Prime Focus Instrument
Package, or PFIP). The VIRUS fiber input heads are connected to the
PFIP and are thus subject to the motions required to both point and
track the telescope. The second question we seek to address is how do the
properties of a VIRUS fiber bundle change over time due to the accumulation of
wear. In order to explore these questions we simulated 10.2 years of
wear (188.7 km of linear travel) on a single VIRUS fiber bundle.
The simulated motion was carried out between February and May, 2011 on
a test rig designed and built for this purpose. The complete details of
the test apparatus are given in \citet{sou10}. In
Figure \ref{setup} we show pictures of the final test configuration
(left two images) and a single frame (right) from a camera placed to
monitor the motion of the optical fibers relative to their protective
conduit.

To briefly summarize the test apparatus, the test platform is moved by 6
actuators, allowing linear motion in X, Y and rotation (rho). The
range of travel is designed to simulate the motion the fibers will
experience from the PFIP. A lower stage provides motion at the bottom
half of the fiber bundle to simulate the motion coming from the
telescope tracking (see Figures 6 and 8 in \citet{sou10}).
Three distinct ``tracks'' were run, simulating a range of observation
tracks that HETDEX will take on the sky. A wide range of track rates
were explored, yet all fall into one of three categories: ``quiescent'',
``observation'', and ``slew''. For the quiescent stage, all motion
is stopped and affords a reference for the tests. In the
observation mode, the fibers are moved at typical telescope
tracking rates (0.77 to 1.3 mm/sec) in order to explore the affect of
fiber motion on our science frames. Lastly, slew mode covers the
fastest rates of motion on the HET, similar to typical
telescope slew rates. A wider range in the rate of motion
was explored in this mode (24.3 to 140 mm/sec), yet as the results
were quantitatively similar, they have been included in the same
designation. In order to reach the goal of 10 years of simulated life
within $\sim 4$ months, the majority of the lifetime tests were run in
the slew mode. As the current VIRUS calibration plan makes use of
the $\sim 90$ second rewind of the telescope for capturing arc and
flat field frames, a clear understanding of the fiber behavior during
the slew mode is also critical to the HETDEX project.

\subsection{METHODS}\label{method}

\begin{figure}
\begin{center}
\begin{tabular}{c}
\includegraphics[scale=0.3,trim=0 0 0 0,clip]{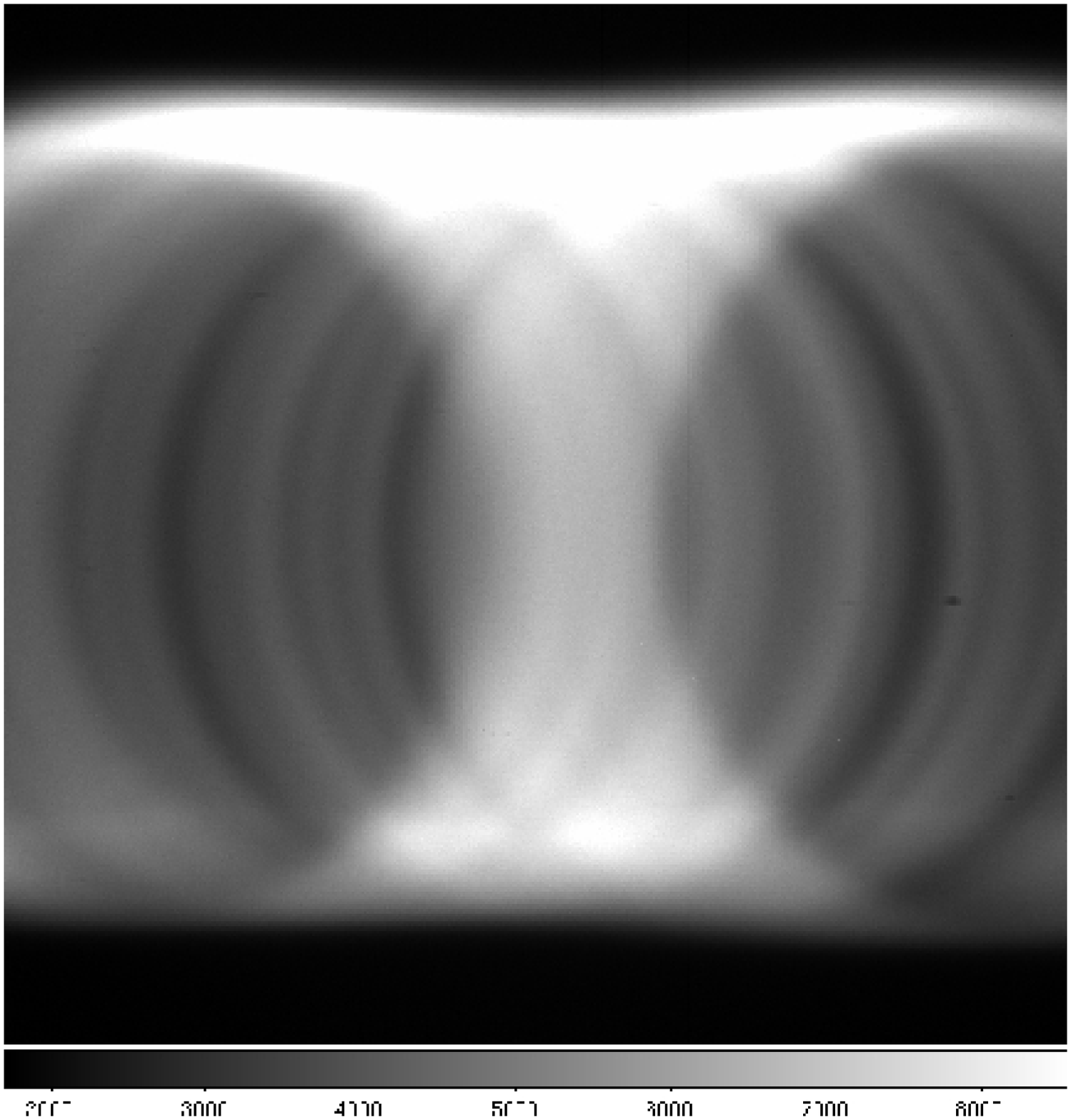}
......
\includegraphics[scale=0.287,trim=0 0 0 0,clip]{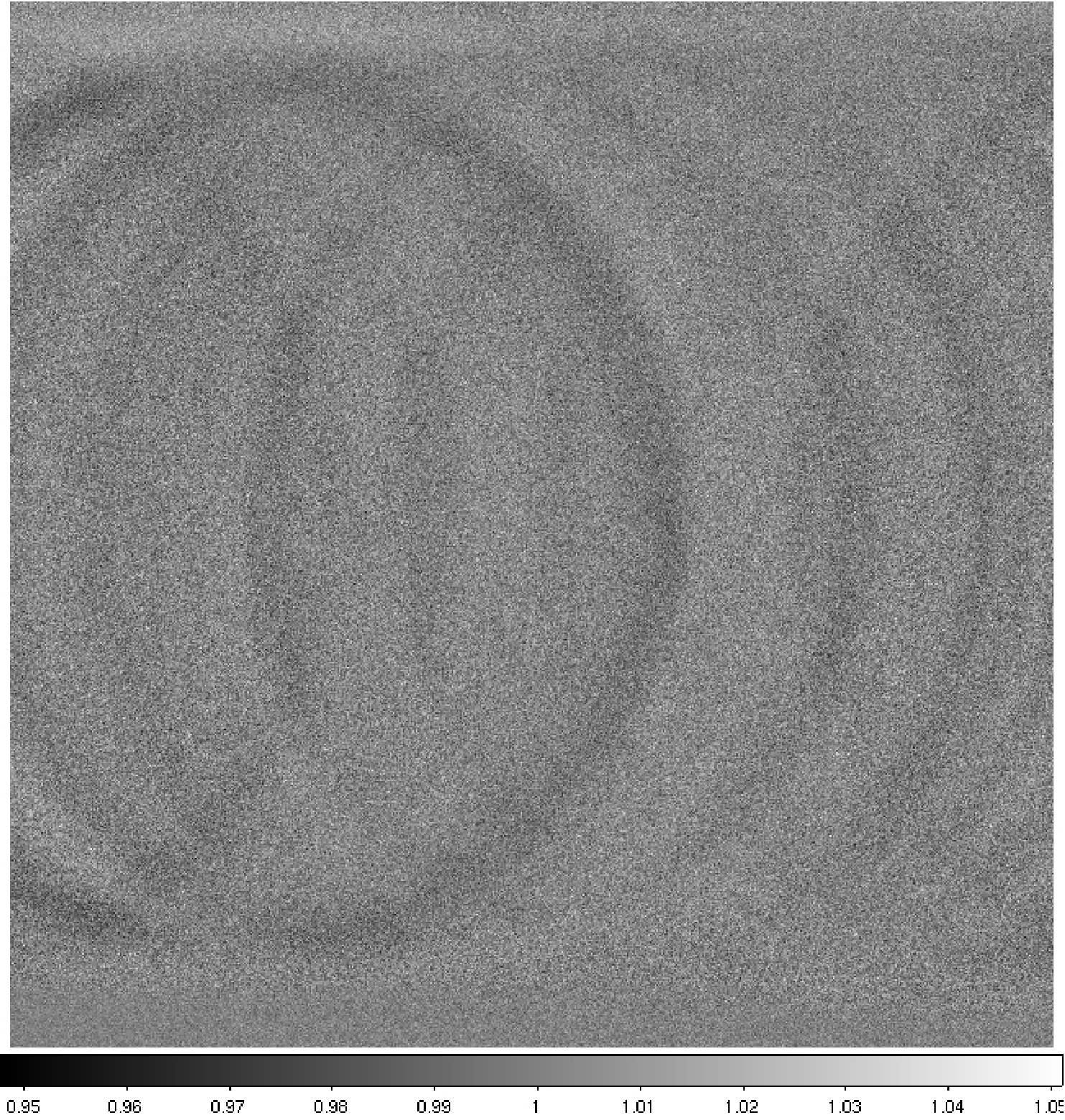}
\end{tabular}
\end{center}
\caption[An example of the divided frames used to quantify the
  severity of FRD seen during the lifetime fiber tests.]{Left: A
  typical far-field frame from the lifetime tests showing the output
  from $\sim 150$ fibers. As described in \S \ref{method},
  frame-by-frame division of neighboring images is used to both
  identify and quantify the transient FRD events seen during the
  lifetime tests. Right: A frame after division by its neighbor. Low
  level, transient FRD events in a few fibers are evident as rings that
  deviate from 1.0. See Figures \ref{dividebad} and \ref{dividegood}
  and the text in \S \ref{method} for further details on how we
  quantify this transient FRD effect.
  \label{frame}}
\end{figure}

For the lifetime tests we wanted to capture potential changes in the
fiber behavior taking place on short timescales (5 to 30 seconds) and
do so over a large number of fibers simultaneously. Yet the tests
described in M08 and outlined in \S \ref{bendtests} are carried out on
individual fibers, with a complete set of transmission and FRD tests
taking $\sim 1$ hour. To achieve the goal of
sampling both a large number of fibers and doing it at a rapid cadence
we took an heuristic approach to the problem, which we describe here.
We start by coupling a collimated beam of light into the input
end of the fibers at a controlled angle. Then, by relying
on the azimuthal scrambling of light
within a fiber, we simulate a full light cone at an input f-ratio of
f/3.65. This approach allows us to couple light into all the fibers
simultaneously, and is conceptually similar to laser-based tests
for FRD \citep{hay08}. The collimated light is filtered with a broad ($\sim
100$~\AA) bandpass filter centered at 500~nm. We then use a
3k~$\times$~3k CCD to image the far-field output of a large set of
fibers ($\sim 150$) simultaneously. These frames 
are taken at a cadence of a single, 1 second exposure every 5 to 30 
seconds, depending on the tests being run. Approximately 250,000
frames were collected over the duration of the lifetime tests. The
left-hand image in Figure \ref{frame} gives an example of a typical
frame taken during testing showing the far-field light of $\sim 150$
fibers. Clearly the light from neighboring fibers is confused 
with the light from other fibers. To overcome this confusion we
employ a differential measurement, wherein frame-by-frame division
is used to detect transient FRD events; by analyzing each divided
frame for deviations from 1.0, and doing so for a large number of
frames over a long baseline, we are able to determine both the
severity of the FRD events and how the events are correlated to the
motion of the test platform. An example of a single divided frame
is shown to the right in Figure \ref{frame}. For each frame,
information on the platform position (X, Y and rho), the lower stage
position, and both temperature and humidity at two locations, near the
input and output ends of the fiber bundle, were recorded. No
measurable affect from humidity was seen, so we do not discuss it
further. Plots of the various parameters against one another (see \S
\ref{motion}) allows us to explore correlations and better understand
the causes of the transient FRD events.

\begin{figure}
\begin{center}
\begin{tabular}{c}
\includegraphics[scale=0.28,trim=0 0 0 0,clip]{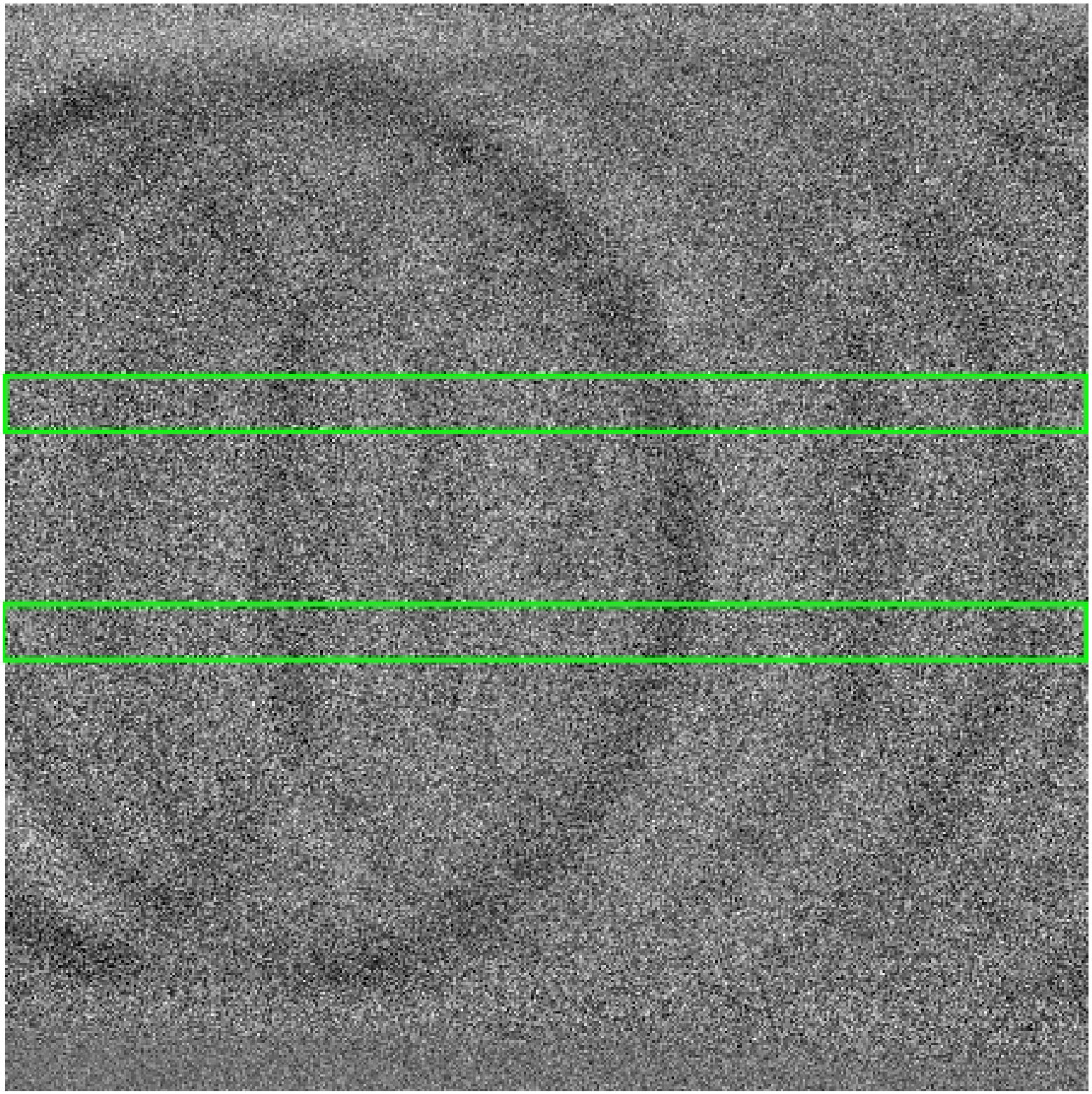}
\includegraphics[scale=0.55,trim=0 5 0 0,clip]{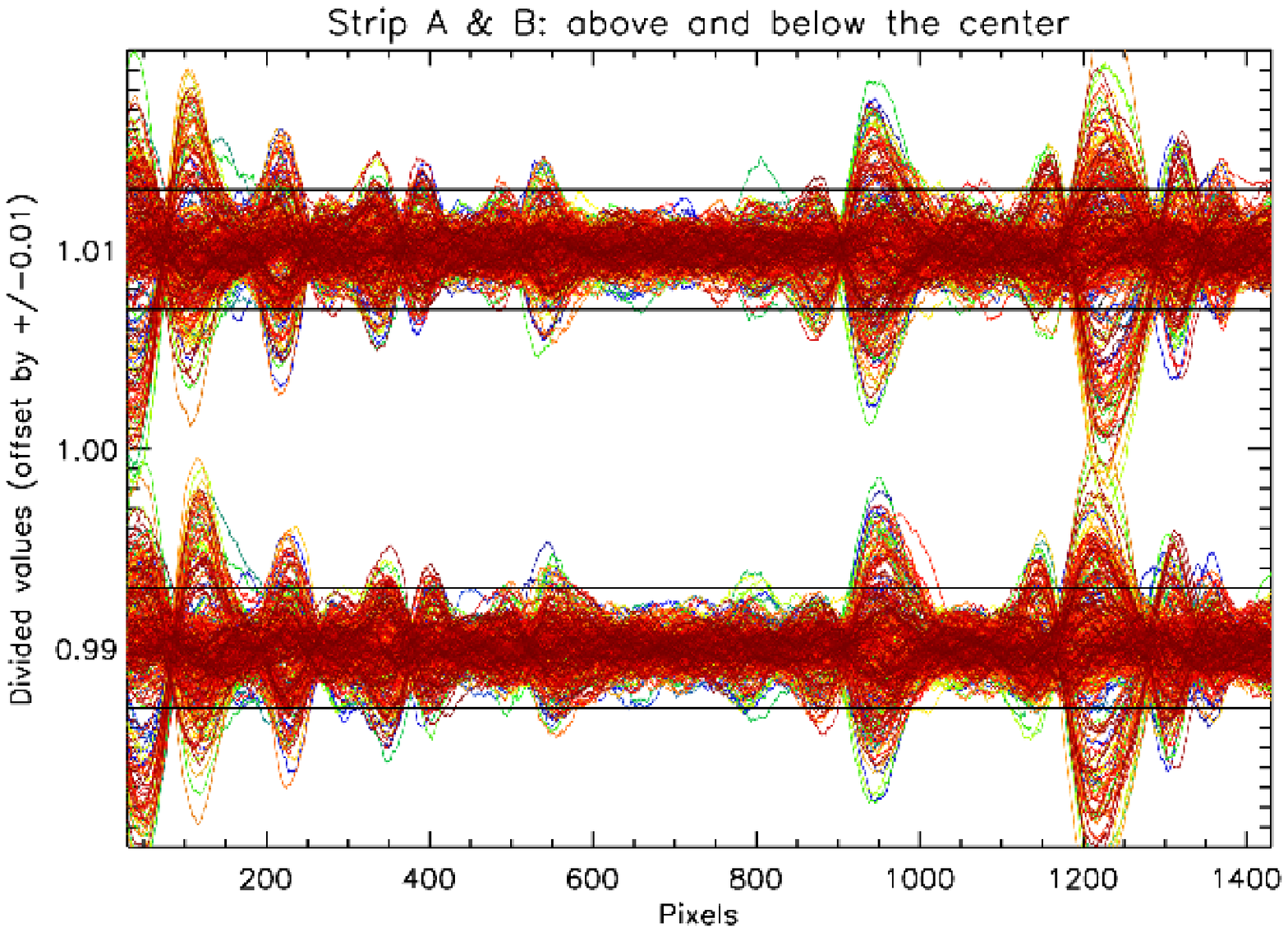}
\end{tabular}
\end{center}
\caption[A divided frame showing evidence for the weak, transient FRD
  events evident throughout the lifetime tests.]{Left: A divided frame
  from a slew rate of travel and showing evidence for weak, transient 
  FRD events in several fibers. Each faint circle, visible only in the
  divided frame, indicates a single fiber that is experiencing some
  level of transient FRD, likely due to a temporary localized shear. The
  horizontal rectangles indicate the regions of the frame that are
  median-combined along the Y-direction, then plotted to quantify the
  severity of FRD. Right: A plot of the median-combined
  frame cross-sections for 400 divided frames. The two regions are
  both centered around 1.00 but have been offset by $\pm 0.01$ in the
  figure for visual clarity. The horizontal black lines indicate the
  $\pm 0.003$ thresholds used to quantify the severity-of-event (SOE)
  parameter as described in the text.
  \label{dividebad}}
\end{figure}

\begin{figure}
\begin{center}
\begin{tabular}{c}
\includegraphics[scale=0.29,trim=0 0 0 0,clip]{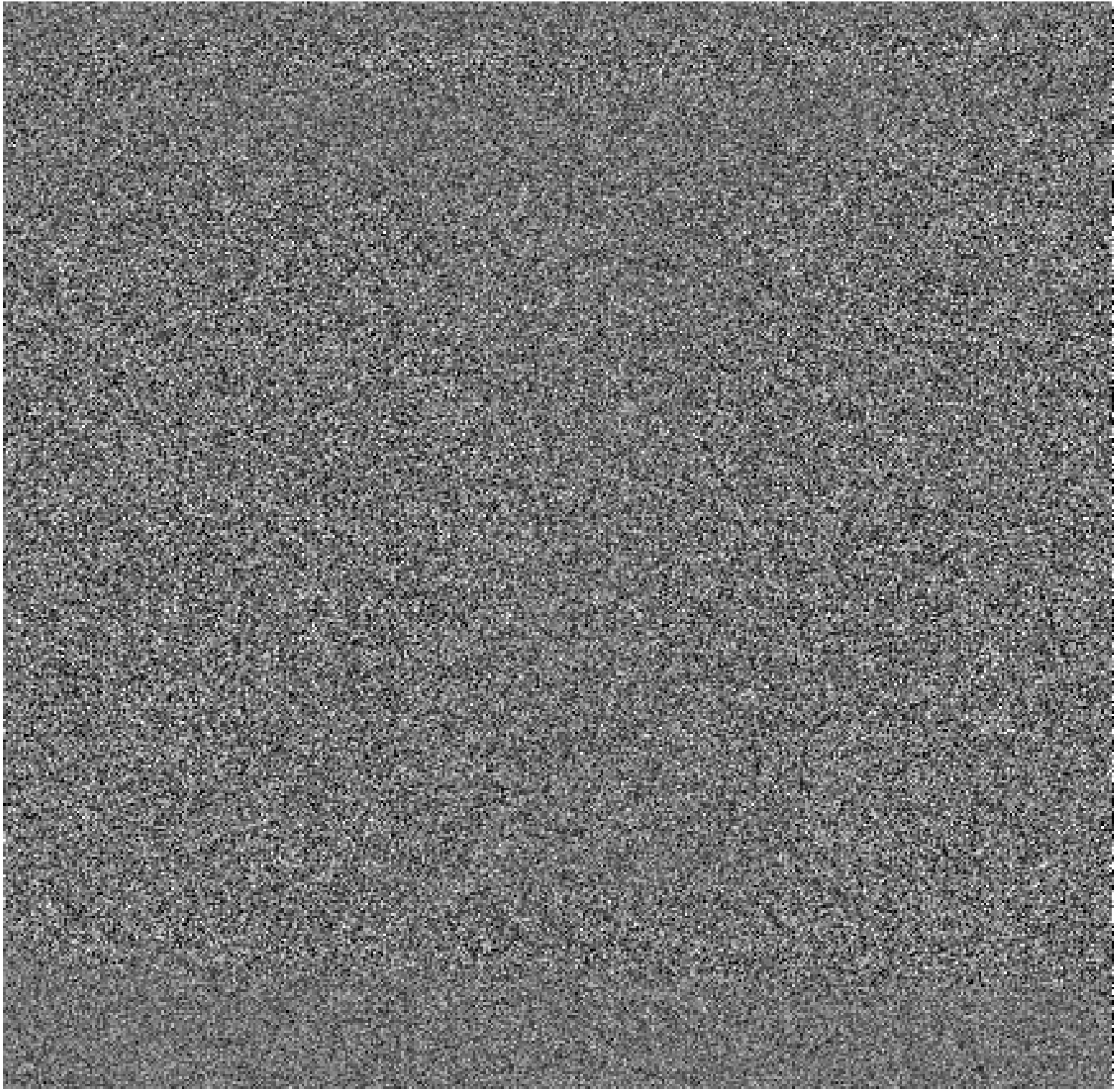}
\includegraphics[scale=0.55,trim=0 5 0 0,clip]{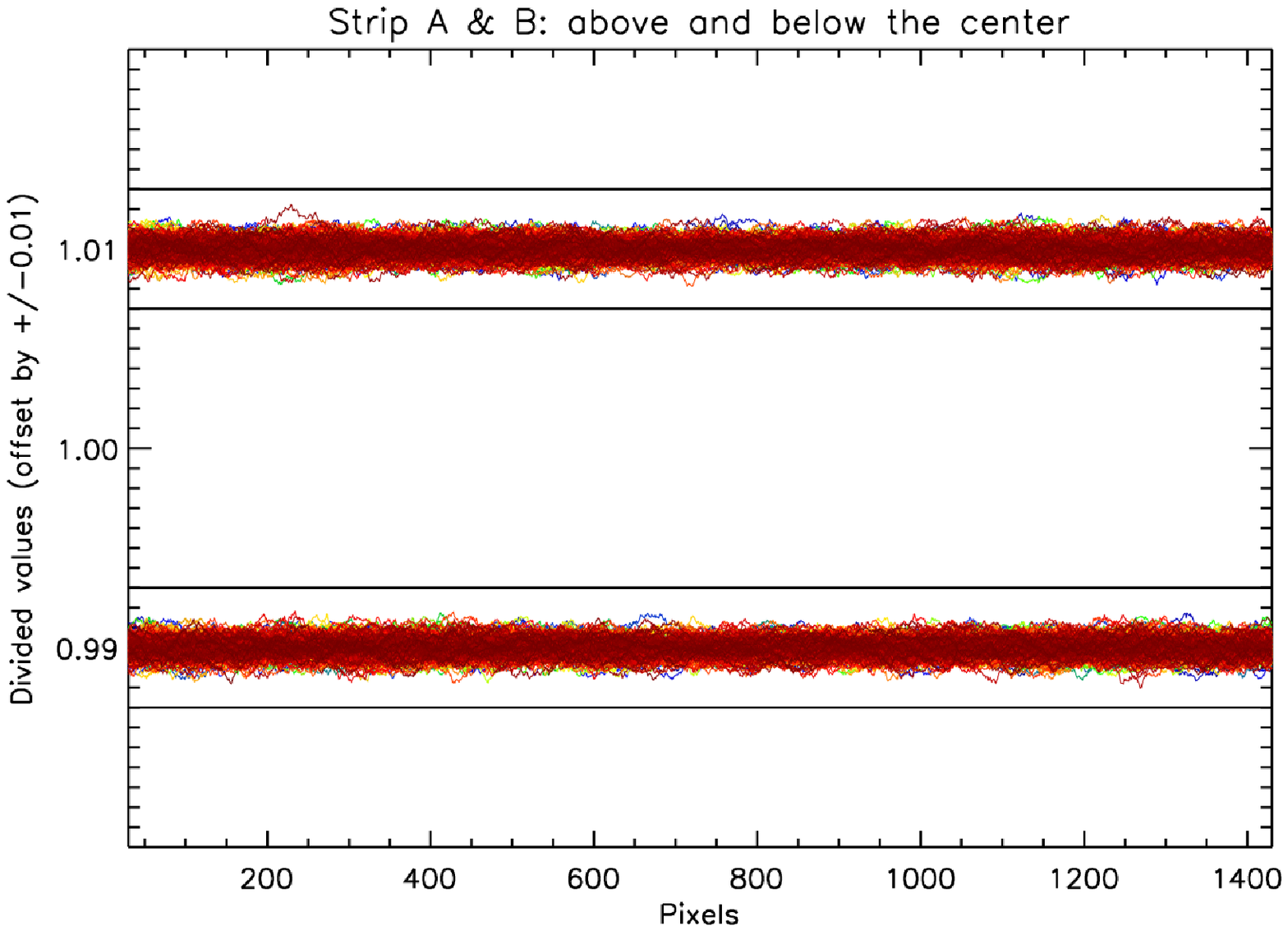}
\end{tabular}
\end{center}
\caption[A divided frame during a quiescent period of the lifetime
  testing showing no evidence for FRD.]{Left: A divided frame, same as
  in Figure \ref{dividebad} but for a quiescent period 
  of the testing. Right: The cross-sections of 400 divided
  frames. None of these 400 frames show any evidence for transient FRD. During
  observation rates, the level of FRD events were extremely low
  and very similar to the frames shown here. See Figures \ref{soe4a}
  and \ref{soe4b} in the Appendix for the SOE results of a track run at
  observation rates. 
  \label{dividegood}}
\end{figure}

In order to quantify the transient FRD events seen in the lifetime
tests, two regions from each set of divided frames are selected for
analysis. These two regions are then median combined along the
Y-direction and plotted as a cross-section through the frame
division. Figures \ref{dividebad} and \ref{dividegood} show examples
of this process for a slew and quiescent period of the testing,
respectively. The left-hand frame in Figures \ref{dividebad} and
\ref{dividegood} show a single divided frame, with the regions
selected for analysis shown as green rectangles in Figure
\ref{dividebad}; the cross-sections of those regions are plotted to
the right. The divided cross-section values are centered around 1.0
yet have been offset in the plot by $\pm 0.01$ for visual
clarity. Note that although we have shown a single, divided frame to
the left in these figures, the cross-sections of 400 divided frames
are plotted in the frames to the right. The evidence for transient FRD
is clearly seen in the cross-sections plotted in Figure \ref{dividebad}.
A threshold is then set for deviations from 1.0. This threshold was
set to $\pm 0.003$ and is plotted as horizontal black lines in the
cross-sectional plots in Figures \ref{dividebad} and \ref{dividegood}.
Although the threshold value is somewhat arbitrary, we are making a
differential measurement and so the exact threshold value chosen is
not important. Once this threshold is set, we simply count up the
number of pixels that fall outside our $\pm 0.003$ threshold in the
cross-sectional cut of a given divided frame. The more counts that
fall outside the threshold, the more severe the FRD is for a given
frame. We will refer to this number as the severity-of-event (SOE)
value. The SOE value is, admittedly, a crude measure in that it lumps
different scenarios under a single value. For example, one could have
low-level FRD in many fibers, or extreme FRD in a single fiber, yet
calculate the same SOE value. However, from inspection of many of the
individual frames, and as evidenced in the cross-section plot of
Figure \ref{dividebad}, the typical case is that the FRD events are
localized to a select number of fibers that get repeatedly affected. 

\subsection{LIFETIME TEST RESULT SUMMARY}\label{results}

The fiber lifetime tests revealed a number of interesting results on
the behavior of optical fibers, both while in motion and the effect of
accumulated wear. We summarize our primary results here, with the
details given in \S \ref{breakage}, \S \ref{motion} and \S
\ref{prepost}:
(1) The fiber bundle went through a period of rapid settling. The outer
conduit stretched by $\sim 4$~cm over the first 4 days. After this
period of settling, the excess conduit length was taken up by an
adjustment made at the output slit (see the right-hand image in Figure
\ref{setup}). Further settling was minimal over the remainder of the
tests, with another 0.3~cm of movement occurring over the subsequent
$\sim 10$ days, and none over the final 3+ months of the tests.
(2) Six fibers broke over the 10+ years of simulated wear. Two of the six
broken fibers were initially weak and were likely already damaged at
the start of the tests. Therefore, the loss of 6 fibers over this time period
is considered an upper-limit.
(3) The motion of fibers at slew rates led to transient FRD events
that tended to cluster around specific locations along a track. The
location of the clustering was dependent on the track being run, with
certain tracks showing a bimodal distribution, and one a trimodal
distribution in the location of the FRD events. We believe these
transient FRD events are due to localized stress and subsequent
mircobending in a select number of fibers.
(4) Transient FRD events for slew rates were common, but at a low
level ($\le 1.0$\%), and drop to effectively zero when the tracks are run
at observation rates.
(5) Although no loss in overall fiber transmission was seen over the lifetime
tests, \emph{clear evidence for a permanent increase in FRD was seen
in all 18 fibers tested both before and after the lifetime tests.}

\subsubsection{FIBER SETTLING AND BREAKAGE}\label{breakage}

\begin{figure}
\begin{center}
\begin{tabular}{c}
\includegraphics[scale=0.24,trim=0 0 0 0,clip]{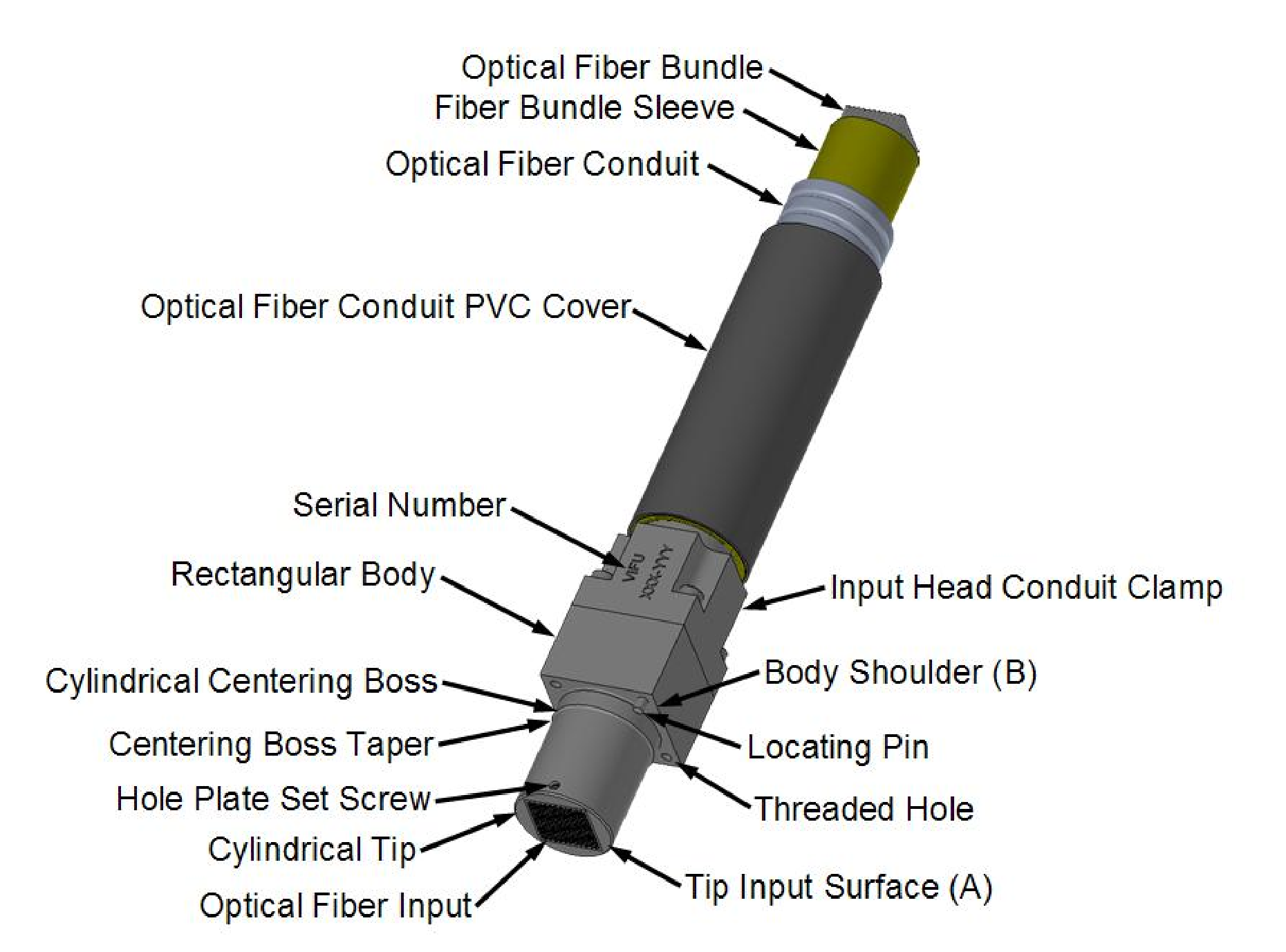}
\includegraphics[scale=0.21,trim=0 0 0 0,clip]{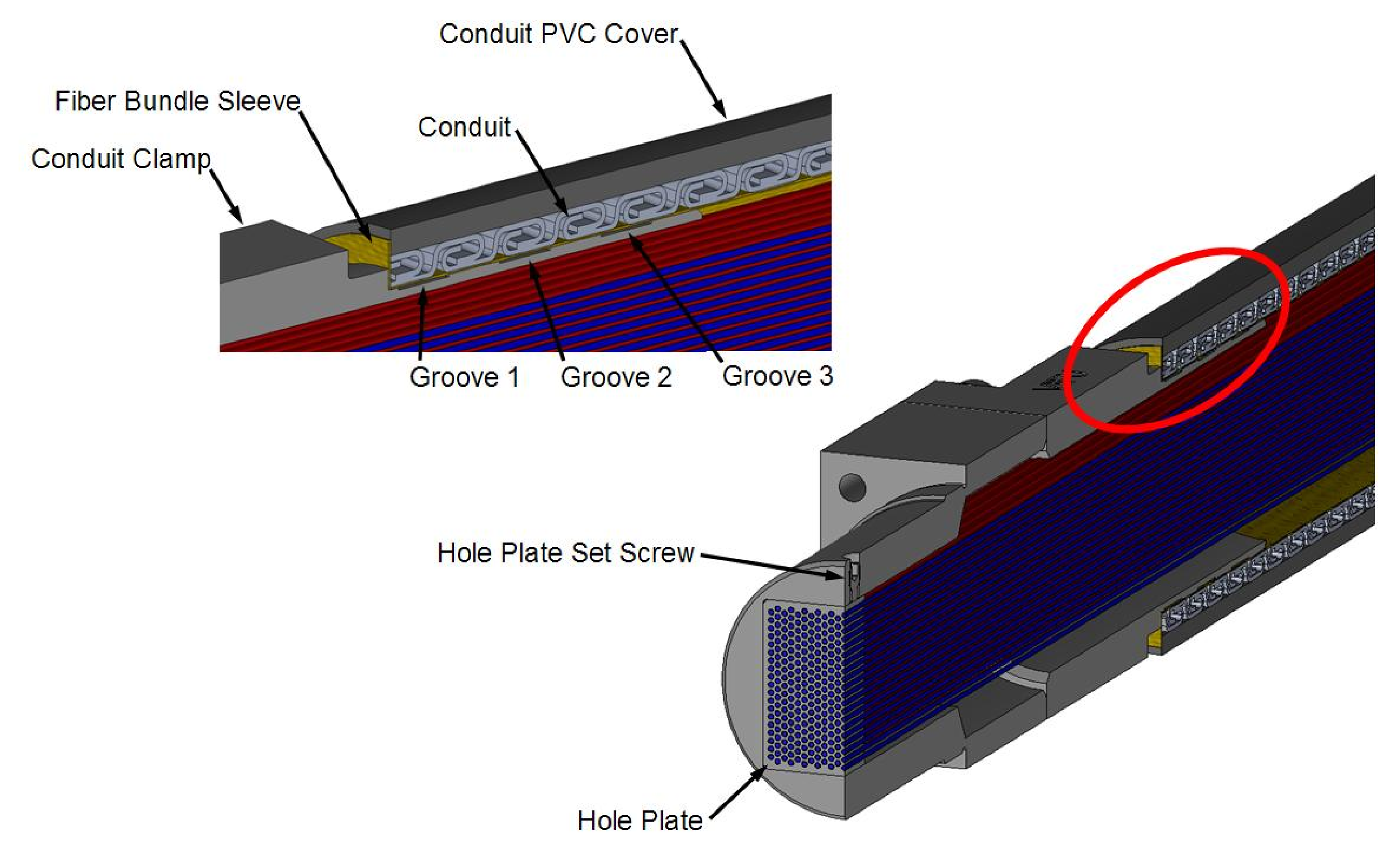}
\end{tabular}
\end{center}
\caption[A schematic of the input head for a VIRUS IFU.]
  {Schematics of the input head for a VIRUS fiber bundle. The 448
  optical fibers are surrounded by a Teflon fiber bundle sleeve to
  reduce friction and minimize wear with the optical fiber conduit. A
  cross-section of the fiber bundle input head is shown to the right.
  \label{head}}
\end{figure}

The motion of the optical fibers relative to their protective conduit was
monitored during the lifetime tests in order to measure the amount of
settling we can expect in the VIRUS fiber bundles. Excessive settling
of the fibers relative to their conduit can lead to fiber tension
and, in extreme cases, fiber breakage. To understand how the
fiber bundle will settle, the position of the fibers relative to their
conduit was monitored at the point where the fibers break out of the
conduit and enter the output slit assembly (see the central image of
Figure 6 in \citet{hil10}). The monitoring was done
with a machine vision camera mounted just above where the fibers enter
the slit assembly. An example of one frame used to monitor fiber
motion is seen in the right-hand image in Figure \ref{setup}.

The fiber bundle experienced some early and rapid settling, with $\sim
4$~cm of fiber being drawn into the conduit as the conduit settled and
lengthened. This settling took place within the first $\sim 4$ days of
testing. The output slit assembly connects to the conduit via an adjustable
tail-piece. We therefore adjusted the tail-piece to take up the excess
conduit length and continued to monitor fiber motion. The fibers
settled another $\sim 0.3$~cm over the next 2 weeks. After that, no
measurable motion of the fibers was observed over the remainder of the
3+ months of the lifetime tests.

Over the duration of the lifetime tests 6 fibers broke. Initially the
test fiber bundle was fabricated with an input head at both
ends. Figure \ref{head} shows a schematic of the input head. This
configuration was initially chosen so that all 448 fibers could be
monitored with a single CCD camera during the tests. However, we later
elected to change this design to better match the final arrangement of
the fibers at the output slit. During the reconfiguration of the fiber
bundle and re-polish, several of the fibers were broken. This didn't
impact our test results, yet may have influenced the number of fibers
that broke  during the lifetime tests, as 2 of the 6 fibers that broke
were noticeably weak in transmission at the start of the tests. We
therefore conclude that the breakage of 6 fibers from wear over 10
years is an upper limit for the VIRUS instrument.

\subsubsection{FIBER MOTION AND TRANSIENT FRD}\label{motion}

\begin{figure}
\begin{center}
\begin{tabular}{c}
\includegraphics[height=90mm,width=160mm]{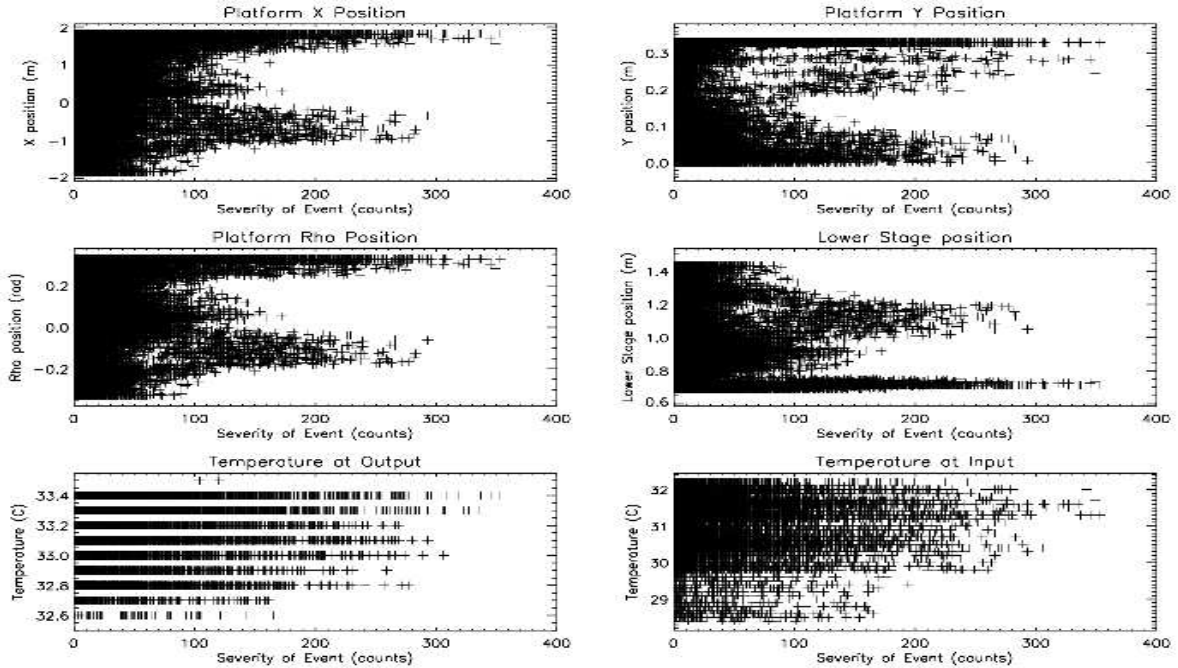}
\end{tabular}
\end{center}
\caption[The severity-of-event parameter plotted against several of
  the parameters measured during the lifetime tests.]{The
  severity-of-event (SOE) value, as described in the text, 
  plotted against various parameters for the Dec 65 track. Several
  thousand frames (18,774), taken over several days, are
  plotted. The top two figures plot the X and Y position of the test
  platform. In the center two figures we plot the rotation (rho) of
  the test platform and the position of the lower stage. The lower two figures
  plot the temperature taken at two locations; lower-left is the
  temperature near the output slit and lower-right plots the
  temperature at the test platform where the light is coupled into the
  fibers. It is clear that FRD increases at the limits of the tracks,
  particularly the Y position. Due to the large number of frames
  plotted, we are able to sample the entire phase space of the 
  test platform many times over many days. This allows us to see
  subtle trends in the data, such as the increase in FRD with
  relatively small increases in temperature. We see a weaker trend
  with temperature for the Dec 38 track (Figure \ref{soe2a}) but not
  for the Dec 60 track (Figure \ref{soe3a}), and so the conclusion
  that FRD increases with temperature is a tentative one.
  \label{soe1a}}
\end{figure}

\begin{figure}
\begin{center}
\begin{tabular}{c}
\includegraphics[height=100mm,width=160mm]{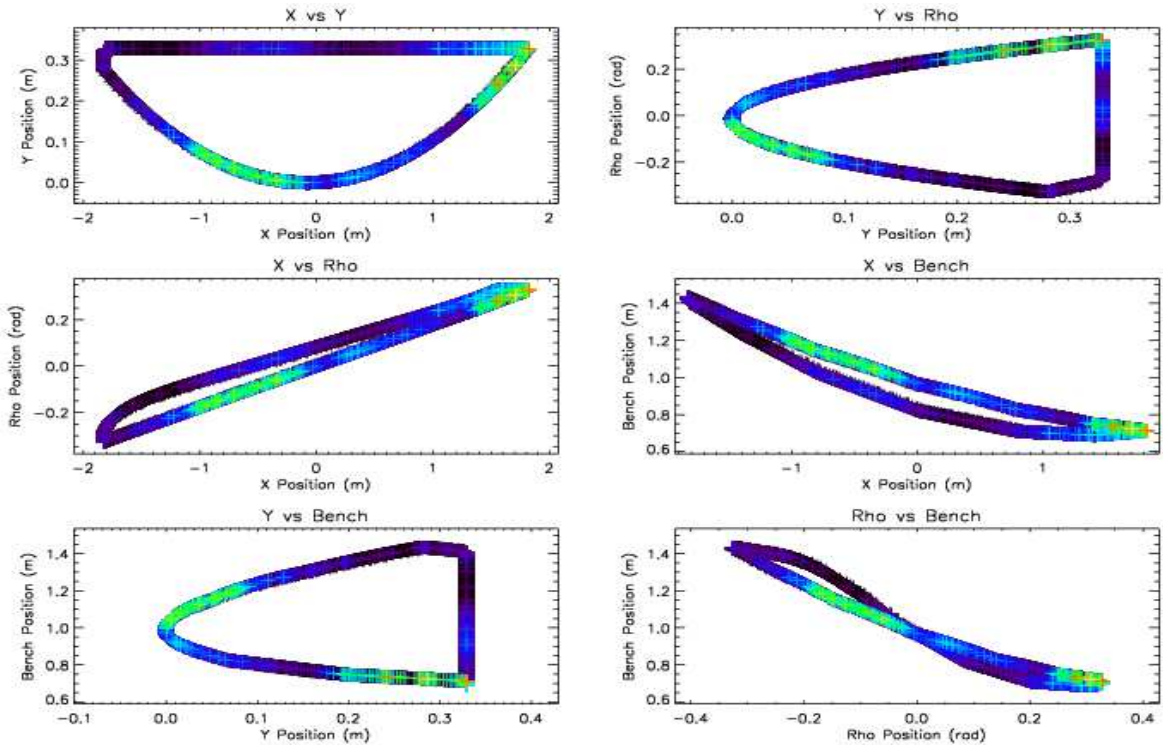}
\end{tabular}
\end{center}
\caption[Various parameters of the lifetime tests plotted against one
  another to show their correlations for the severity-of-event
  values.]{The 4 position parameters (X, Y, rho, lower stage) plotted 
  against one another in order to better understand the correlations
  between the positions of the test platform and the transient FRD
  seen in the lifetime tests. The SOE value is indicated by
  color, ranging from zero (plotted as black) to the highest SOE values
  (plotted as light green to red). There are two primary sections
  along the path (both turnaround points for the Y 
  motion) where FRD events are seen to increase. Figures \ref{soe2b},
  \ref{soe4b} and \ref{soe3b} plot the results for 3 different
  tracks and rates in the same manner as shown here.
  \label{soe1b}}
\end{figure}

With an SOE value attached to each set of divided frames (as described
in \S \ref{method}) we can plot
the various parameters, both against the SOE values and against one
another. This proves extremely informative as we are then able to
determine at which positions the transient FRD events are most common, and, with
thousands of repeat measurements, begin to see subtle trends in the
data. Figures \ref{soe1a} and \ref{soe1b} show the effects of motion on the
fibers during the lifetime tests. In Figure \ref{soe1a}, the SOE value
is plotted against 6 different parameters in order to map out where
the transient FRD events are the most severe. The top two rows in
Figure \ref{soe1a} plot the position of the test platform (X, Y, rho)
and lower stage against the SOE number. The lower row plots temperature at the
output slit (left) and the input head (right). Nearly 20,000 frames
are plotted here, all taken in slew mode and spanning just over
5 days. In Figure \ref{soe1b} the same data set is shown, only with
the various motion parameters plotted against one another in order to show
correlations. The color indicates the SOE value, with black being an
SOE value of 0 and red being the highest value seen in Figure
\ref{soe1a}. For this track it is clear that the FRD events that drive
the SOE number to higher values occur at two locations along the
track. In the Appendix (Figures \ref{soe2a} to \ref{soe3b}) we show
similar figures for a subset of different tracks and rates run during
the lifetime tests, including tests run at observation rates (Figures
\ref{soe4a} and \ref{soe4b}).

In all 3 slew tracks, the high SOE values ranged from $\sim
300$ to over 400, yet the transient FRD effects dropped to nearly
zero for the same tracks when run in the observation mode. Clearly the
speed at which the fiber bundle is moving, rather than the track
itself, is what drives the transient FRD events. This begs the
question, if the transient FRD events are caused by short-term,
localized stress, as we hypothesize, then why do the fibers that
exhibit FRD at certain locations along the track not exhibit
that same FRD at observation rates? A difference in frame
sampling cannot explain this, as we have compared frames over a range
of different time cadences (e.g. a frame is divided by frames other
than its nearest neighbor).

We believe the difference between the slew and observation rates is
due to a difference in settling time; at slew rates the fibers do not
have a chance to smoothly settle. This leads to regions of temporary,
localized stress. Based on the cross-section plot in Figure
\ref{dividebad} it is clear that a relatively small number of fibers
are being affected repeatedly. Along certain points in the track these
fibers experience localized pressure as they roll over one another. In
the slew track rate, the fibers don't have time to alleviate the
pressure between the fibers, leading to low-level, transient FRD. In
the observation mode, the rate is slow enough to allow proper fiber
settling and the localized points of stress are avoided.

\subsubsection{BEFORE AND AFTER TESTS}\label{prepost}

\begin{figure}
\begin{center}
\begin{tabular}{c}
\includegraphics[scale=0.47,trim=0 0 0 0,clip]{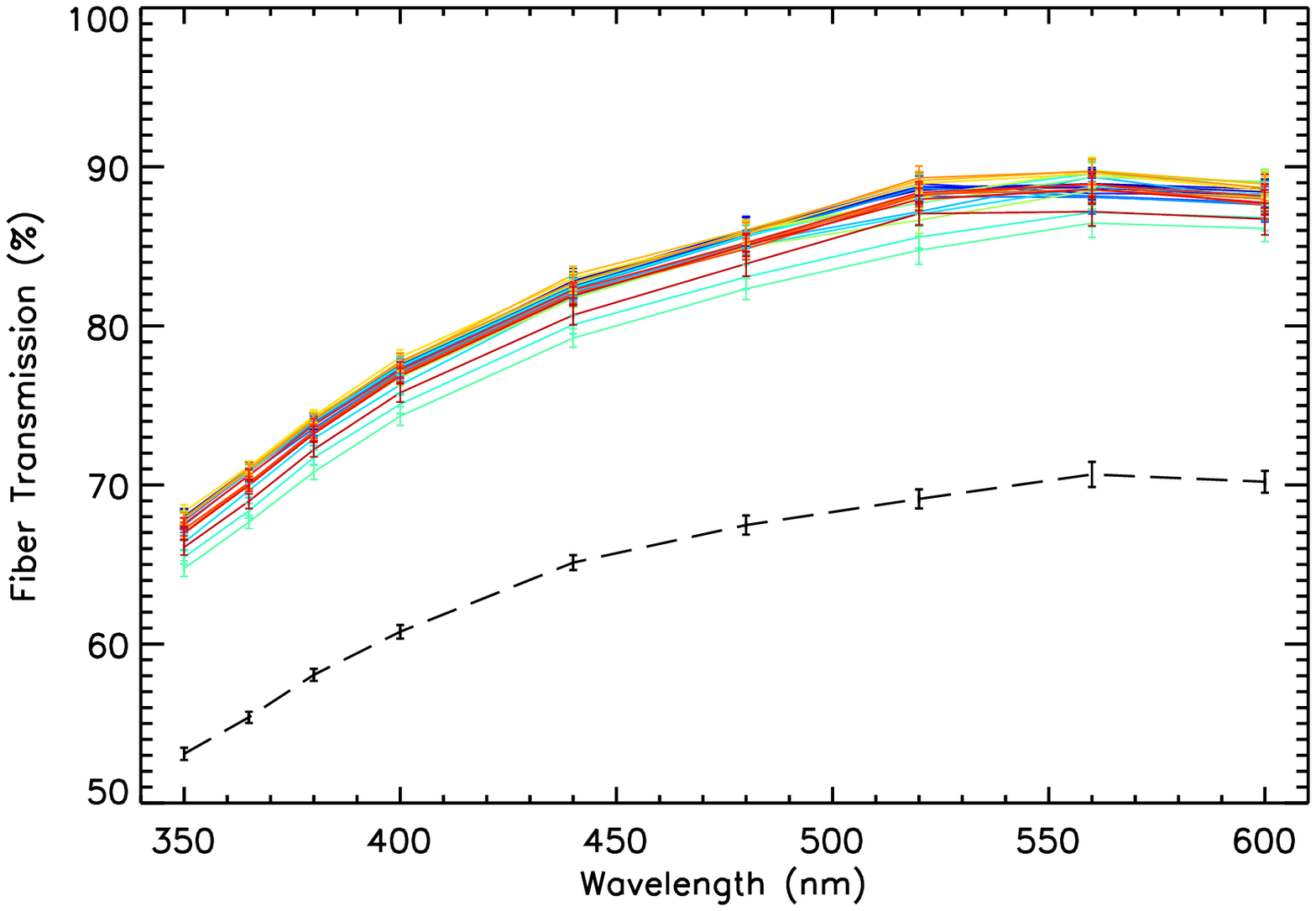}
\includegraphics[scale=0.47,trim=0 0 0 0,clip]{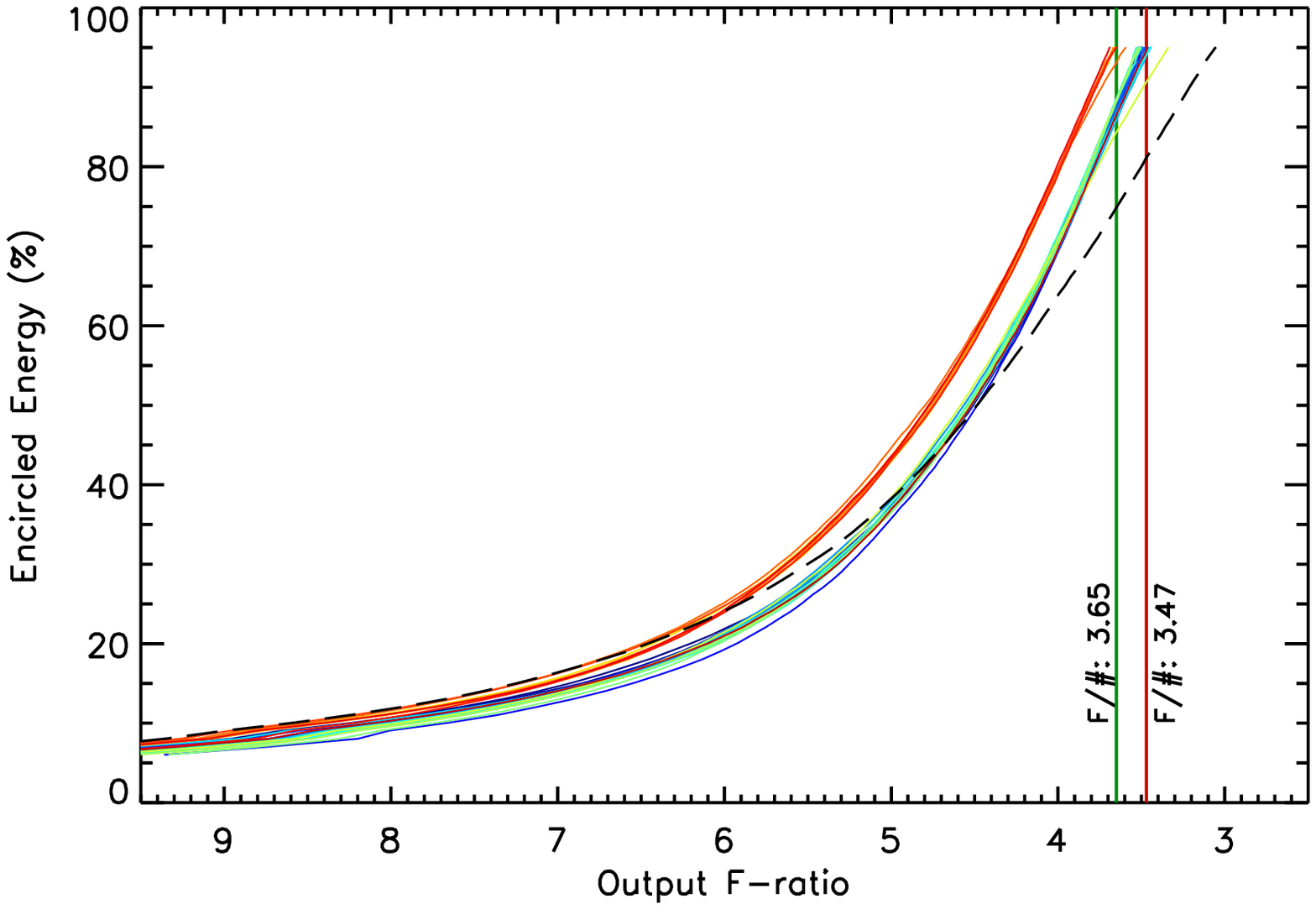}
\end{tabular}
\end{center}
\caption[Fiber transmission values for the lifetime fiber bundle taken
  both before and after the lifetime tests.]{Left: Transmission values
  at 9 discrete wavelengths ($\sim 10$~nm FWHM) for a set of 18
  fibers tested before the lifetime tests. The baseline frames needed
  to determine fiber transmission were corrupted for 7 of the 25
  fibers tested before the lifetime tests. Therefore, only FRD
  measurements could be made for these fibers. The transmission tests
  were conducted without cover-plates or indexing matching gel. Based
  on results in M08, we expect the transmission values to increase by
  $\sim 4$\%\ at all wavelengths once the cover-plates and
  index-matching gel are installed.
  Right: EE vs. f/out, similar to Figure \ref{frdvee}, but  
  for all 25 fibers tested before the lifetime tests. Even without the
  installation of the cover-plates, and the generally low quality of
  final polish on the test fiber bundle, 22 of the 25 fibers meet the
  specification for VIRUS (f/3.47 at 95\%\ EE). The fiber showing
  low transmission (dashed line) in the left figure clearly exhibits
  the worst FRD of all fibers tested. Note the bimodal distribution in
  the FRD values. The input head had a large section near
  one side that showed a visibly worse polish. This bimodal
  distribution is directly correlated with the region of poor polish
  and the location of the fibers at the input head, and reveals the
  influence that fiber end preparation has on FRD. In Figure
  \ref{frdafter}a we plot a subset of these fibers retested after the
  lifetime tests. Interestingly, the affect poor fiber polish has on FRD
  has been washed out by the general increase in FRD due to the
  accumulated wear of the lifetime tests.
  \label{before}}
\end{figure}

\begin{figure}
\begin{center}
\begin{tabular}{c}
\includegraphics[scale=0.8,trim=0 0 0 0,clip]{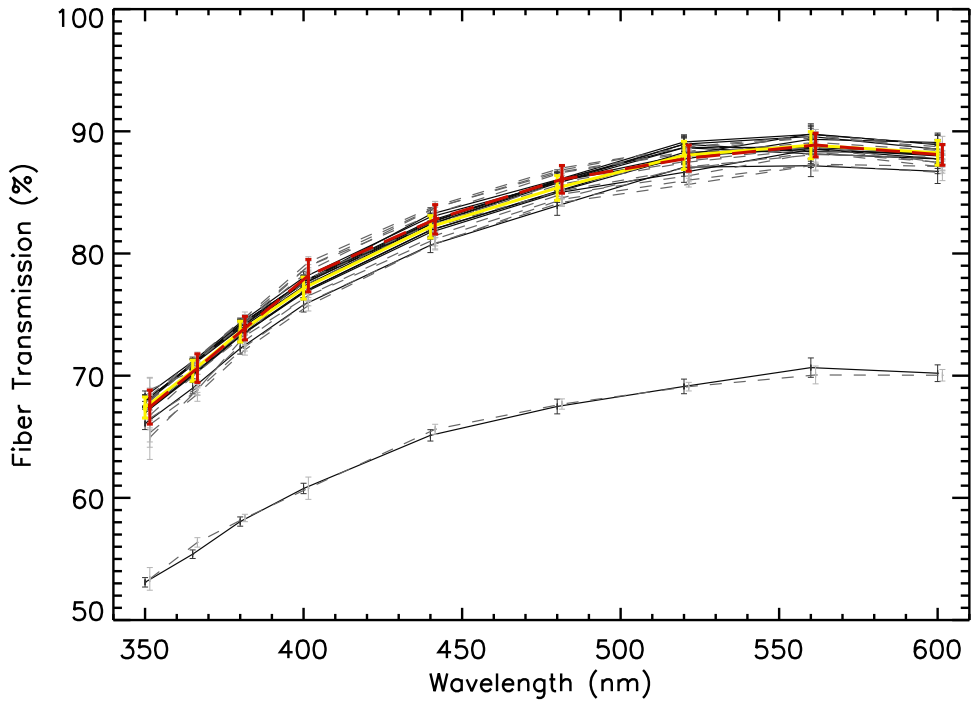}
\end{tabular}
\end{center}
\caption[A comparison of the average fiber transmission values from
  before and after the lifetime fiber tests.]
  {A comparison of the transmission of 18 fibers tested both before
  (solid, black lines) and after (dashed, gray lines) the lifetime
  tests. The uncertainties shown indicate the full scatter from 7 repeat
  measurements over a period of $\sim 1$ hour. After suppressing the 
  clearly weak fiber, the average of the remaining 17 fibers is
  calculated at each wavelength and plotted as the heavy solid yellow
  line and heavy dashed red line for the before and after
  transmission values, respectively. The uncertainties plotted for the
  averages are the standard deviation of the 17 fiber values. No
  change in the overall fiber transmission is seen from the 10+ years
  of simulated wear
  \label{transafter}}
\end{figure}

Before the lifetime tests were carried out, the test fiber bundle was
characterized for both transmission and FRD. Twenty-five fibers were tested
before the lifetime tests and 18 of those same fibers were re-tested after.
In Figure \ref{before} we plot both the transmission and FRD results
for the 25 fibers tested before the lifetime tests. As we wanted to observe
unmitigated fiber FRD, the input and output lens and index-matching
gel were not installed. As previously shown (compare
Figures 10 and 11 in M08) the index-matching gel and cover-plates are
expected to boost the transmission values by $\sim 4$\% at all
wavelengths. Also, the input end surface polish on the test fiber
bundle was below specification. Therefore, we expect the FRD and
transmission values shown in Figure \ref{before} to improve for the
final VIRUS science fiber bundles.

\begin{figure}
\begin{center}
\begin{tabular}{c}
\includegraphics[scale=1.05,trim=20 10 0 0,clip]{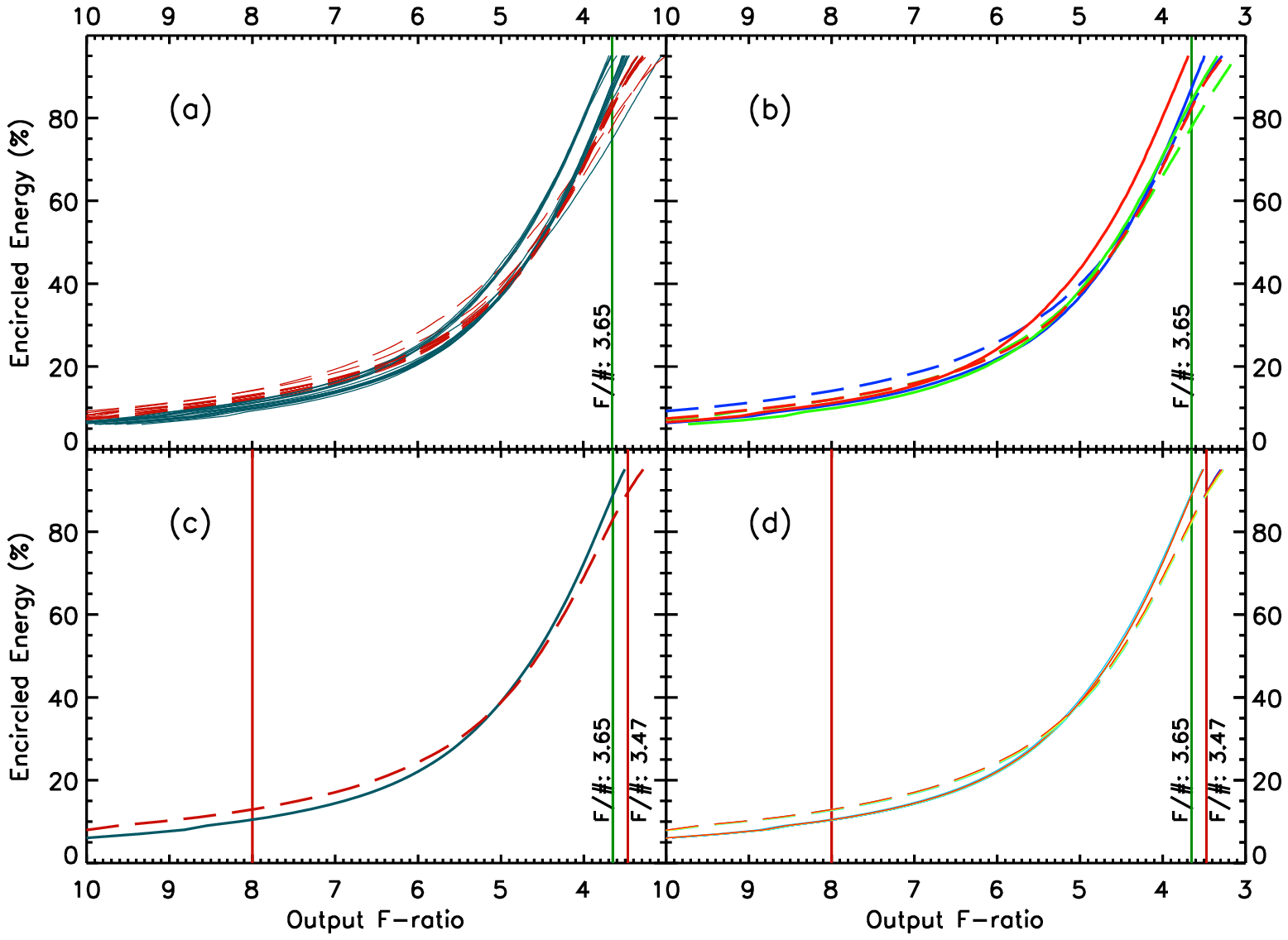}
\end{tabular}
\end{center}
\caption[A comparison of the FRD values taken both before and after
  the lifetime tests showing evidence for an increase in FRD stemming
  from the accumulation of wear.]
  {A figure comparing the FRD measured from before and after the 
  lifetime tests. As we find no dependence of FRD on wavelength (see plot
  'd') we show only the results at 480~nm in plots 'a', 'b' and 'c'.
  (a) FRD values for all 25 fibers tested before the lifetime tests
  are plotted as solid blue lines. The 18 fibers tested after
  the lifetime tests are plotted as dashed red lines. The increase
  in FRD from the accumulated wear of the lifetime tests is clearly
  evident. \emph{Note that the bimodal distribution of FRD 
  seen before the lifetime tests (as discussed in Figure
  \ref{before}) has been washed out by the increase in FRD stemming
  from the accumulation of wear.} (b) The FRD of 3 fibers selected for closer
  examination. Each color represents a different fiber, with solid
  lines indicating the values measured before the tests, and dashed lines
  indicating the values measured after the lifetime tests. Comparing
  the before and after values, all three fibers show increased FRD,
  yet the way the FRD manifests is different. For the red fiber, the
  majority of FRD occurs in the outer halo. This is the opposite for
  the blue fiber, where the predominant increase in FRD is seen in the
  central obscuration. The light green line shows a balance between
  light lost into both the center and the outer halo. (c) The average
  of 17 fibers tested before (solid blue) and after (dashed red) the lifetime
  tests. The inner vertical line shows the approximate extent of the
  shadow of the PFIP. Light within this region (i.e. higher f/out numbers)
  is lost onto the back of the CCD. A line indicating the f/3.47
  acceptance f-ratio for VIRUS is also shown. The increase in FRD is
  clear. (d) A figure plotting the average (as in (c)) of all 9 
  wavelengths tested both before (solid) and after (dashed) the
  lifetime tests. The colors run from blue (350 nm) to red (600 nm);
  with no evidence of a wavelength dependence on FRD, either before or
  after the lifetime tests, the lines are all on top of one another
  and, therefore, not visibly discernible.
  \label{frdafter}}
\end{figure}

Figure \ref{transafter} plots fiber transmission from both before and
after the lifetime tests. After rejecting the fiber that shows very low
transmission, we calculate and plot the average of the before (solid,
yellow) and after (dashed, red) data. \emph{We find no measurable change in the
  transmission of the VIRUS fibers, at any wavelength, over the 10+
  years of simulated wear.} Because we count all photons that pass
through the fiber, and are not concerned with the effects of FRD,
the transmission values before and after the test agree remarkably
well, despite the clear increase in FRD seen.

In Figure \ref{frdafter} we plot the primary results of our FRD
measurements taken before and after the lifetime tests. A second major
finding of the lifetime tests is that \emph{we see clear evidence for
  increased FRD in all 18 fibers re-tested at the conclusion of the
  lifetime tests.} Although the total transmission of the fibers
was unaffected, an increased level of light was seen in both the central
obscuration and outer halo. This is best seen in Figure
\ref{frdafter}c where we have plotted the average of 17 of the 18 fibers
tested both before and after the tests; the fiber exhibiting extremely
poor FRD was withheld. The FRD curve taken after the lifetime
tests (dashed, red line) clearly shows more light lost to the central
obscuration \emph{and} the outer halo. The increase in FRD was seen at
the same level at all 9 wavelengths tested, as seen in Figure
\ref{frdafter}d. The before (solid) and after (dashed) curves for all
9 wavelengths are overplotted from blue to red and are nearly
identical. We refer the reader to the caption of Figure \ref{frdafter}
for a complete description of the plots. Although the cause of this
rise in FRD is not known with certainty, we speculate it is due to an
increase in microfractures stemming from repetition of motion of the optical
fibers. This type of FRD stands in sharp contrast to the transient FRD
events that are driven by localized shear and that typically occur at
the limits of the track range.

We can use these FRD measurements from before and after the lifetime
tests to make a crude estimate of the rate of transmission loss we expect
to see in the VIRUS instrument due to increased FRD. This estimate
necessarily makes a number of simplifying assumptions. First, we
assume a circular central obscuration that does not change. This is a
simplification for two reasons. One, the pupil of the instrument has
both a wavelength and in-slit fiber position dependence; the location
of the CCD relative to the central obscuration within the instrument
changes \citep{lee10}. Two, the illumination pattern of the HET is also
variable and changes over a given track, so the light coupling into a
fiber varies. Setting aside these complicating factors, we can estimate the
percentage of light lost both into the central obscuration, where light
will be blocked by the CCD, and outside the acceptance f-ratio of
VIRUS. We estimate the light lost to the central obscuration at 2.7\%\ over
a 10 year life span. The loss of light outside the acceptance f-ratio
is greater and measures 5.8\%. Therefore, the total loss due to increased
FRD over a 10 year lifespan is 8.5\%. If we make a further simplifying
assumption of a linear increase in FRD over that time, this gives a
through-the-instrument transmission loss rate of 0.85\% per year due to
increasing FRD effects.

\section{MITCHELL SPECTROGRAPH FIBER BUNDLE DEGRADATION}\label{mspec}

\begin{figure}
\begin{center}
\begin{tabular}{c}
\includegraphics[scale=0.85,trim=0 0 0 70,clip]{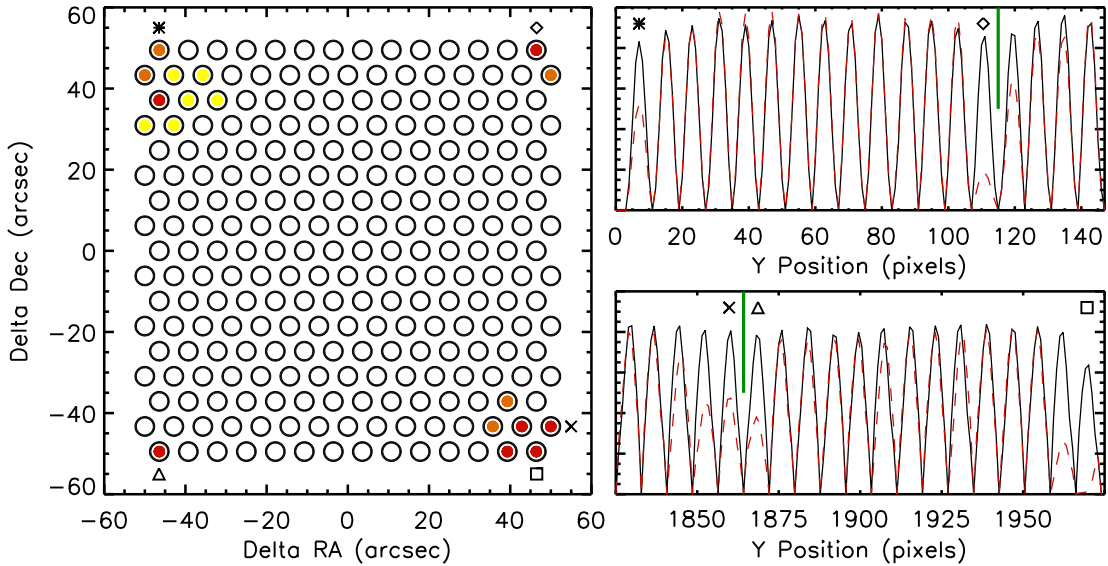}
\end{tabular}
\end{center}
\caption[On-sky map of the VP2 optical fibers that are showing
  evidence for severe FRD due to localized stress at the IFU input
  head.]{Left: An on-sky map of the VP2 fibers showing those fibers 
  most affected by FRD due to localized stress. The fibers are
  over-sized slightly in the figure for display purposes. The
  colors indicate the approximate severity of 
  the FRD and subsequent loss of transmission. Yellow denotes a 5\%\ to
  20\%\ loss, orange shows a 20\%\ to 40\%\ loss, and red indicates
  transmission loss $\ge 40$\%. Right: Cross-sectional profiles of
  fiber transmission taken on October 2008 (solid, black line) and
  March 2011 (dashed, red line) from flat-field frames at $\sim
  4500$~\AA. Tests at other wavelengths return very similar results. The
  drop in transmission of individual fibers is clearly 
  evident. The symbols above specific fibers reference where they are
  located on the IFU head in the left-hand figure. The vertical
  (green) lines locate the end of a row of fibers. Note that these
  cross-sectional profiles include all losses through the instrument. This
  includes central obscuration loss (onto the back of the CCD), outer
  halo loss (outside the acceptance f-ratio of the instrument), and,
  for those fibers with large losses, a suspected breakdown of total
  internal reflection within the fiber.
  \label{vp2}}
\end{figure}

\begin{figure}
\begin{center}
\begin{tabular}{c}
\includegraphics[scale=0.85]{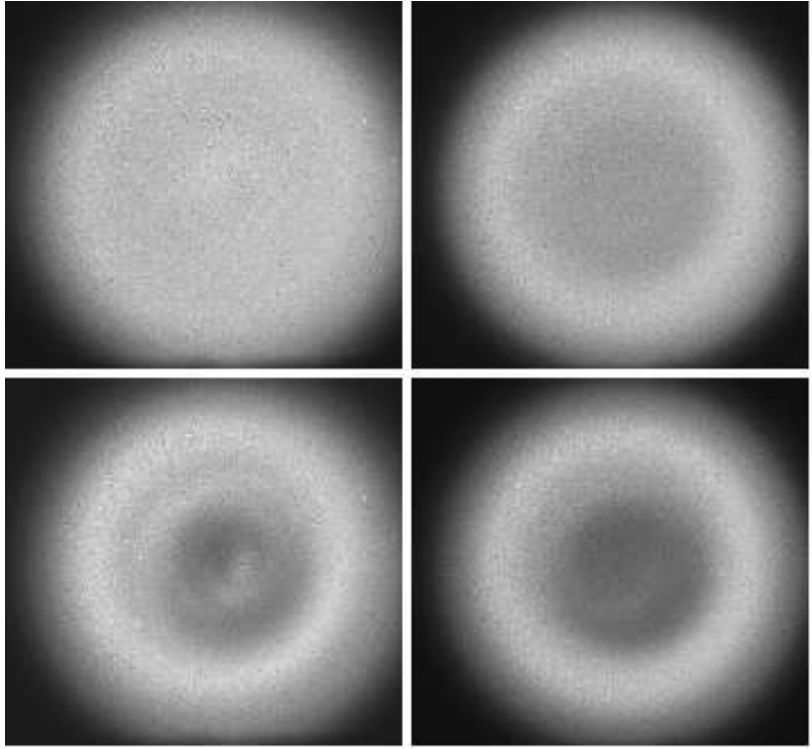}
\end{tabular}
\end{center}
\caption[Far-field images showing the results from twist tests used to
  confirm the source of fiber transmission loss seen in VP2.]
  {Far-field images of a single fiber in VP2 affected by FRD stemming
  from localized shear at the input head of the IFU. The fiber shown
  here is denoted by the 
  triangle in Figure \ref{vp2}. The four frames shown come from
  different levels of twist placed into the fiber bundle. Even in the
  most extreme case (upper-left), where the central obscuration is
  entirely filled, the twist applied was a single ``figure 8'' put
  into the bundle $\sim 10$~m away from the input head. In the
  lower-right corner the bundle has been fully relaxed and the FRD is 
  substantially reduced.
  \label{vp2ff}}
\end{figure}

The Mitchell Spectrograph (formally VIRUS-P) \citep{hil08b} is an
integral-field spectrograph that has been in use on the Harlan
J. Smith 2.7~m telescope at McDonald Observatory for the past 6
years. The current fiber bundle, VP2, has been the primary science
bundle from Spring, 2008 to the present. VP2 is composed of 246 optical
fibers, each with a 200 $\mu$m core diameter. Complete
transmission and FRD tests for VP2 can be found in M08. Here, we report
on a drop in overall fiber transmission in a select number of fibers seen
in VP2. This transmission drop is caused by severe FRD due to a
localized shear in the affected fibers that occurs near the input head
of the fiber bundle, which is similar to the design shown in Figure \ref{head}.
Figure \ref{vp2} plots the on-sky positions of all the fibers in VP2.
When the fiber bundle was first commissioned in Spring,
2008, all 246 of the optical fibers were within specification and
performing well on sky \citep{mur08}. However, over time we have seen a
drop in the transmission of the 18 fibers highlighted in Figure
\ref{vp2}. The colors in the left-hand plot of Figure
\ref{vp2} indicate the severity of the decrease as described in the
figure caption. This drop in transmission is seen clearly in the flat field
cross-sectional profiles, shown to the right in Figure \ref{vp2}, and
can reach $\ge 85$\%\ in the worst cases.

The drop in fiber transmission was first noticed during an observing
run in December 2008. VP2 is 22~m long, yet the length needed for the
Mitchell Spectrograph is just over 4~m. As the Mitchell Spectrograph
is a prototype for the VIRUS instrument, the extra fiber length was
necessary for engineering runs on the HET during the design of VIRUS. When
the Mitchell Spectrograph is in operation on the 2.7~m telescope, the
extra length of VP2 was being taken up by a D~$\sim 1$~m coil attached
to the side of the instrument. With the discovery of the drop in
transmission of the corner fibers in VP2, we speculated this was due
to accumulated stress in the fibers and subsequent FRD; due to the clear
trend in the location of the affected fibers, it was hypothesized that
we were seeing an explosive form of FRD in select fibers stemming from
accumulated twists that propagate the length of the fiber bundle and
terminate at the input head. As the fibers are glued in place at the
input head, the twists can no longer propagate and a shear develops between a
fiber and its point of termination. Fibers at the outer edges of the
input head take up the majority of this transmitted shear and thus
exhibit the highest FRD. The fiber bundle was uncoiled and
shaken out during an observing run in February 2009, and the transmission
of all affected fibers returned to normal.

VP2 was taken back to the lab where twist tests were conducted on
the fiber test bench. Figure \ref{vp2ff} shows 4 far-field images of
a single corner fiber (denoted with a triangle in Figure \ref{vp2}),
with various levels of twist placed into the fiber bundle. From these
tests we confirm that the loss in transmission was an extreme
form of FRD due to localized shear at the input head, with light
being lost into the central obscuration and outer halo. Surprisingly, the
test twists were placed into the fiber bundle \emph{nearly 10~m away
from the input head,} and reveals how far fibers can propagate
stress before manifesting its results. It is very likely that, in the
most extreme cases, where overall transmission loss exceeds 40\%, the
localized shear is causing loss of total internal reflection and
thus significant light loss into the fiber cladding. Regrettably,
testing for loss of total internal reflection was not done while VP2
was in the lab, so this hypothesis remains untested. We also
note that, like the FRD seen in the lifetime tests, we see no evidence
for a dependence on wavelength in the loss of transmission due to
localized shear. In \citet{pop07} they see a trend
of increasing FRD with wavelength over the 450--700~nm range of their
tests. However, this trend is clear in their data at the slower f-ratios
of their tests (e.g. f/in $\ge 10$), but is not readily apparent in
their results at f/in~$\le$~5, which is the range for our tests.

Since the discovery of this FRD effect, the fiber bundle was re-routed
on the telescope to remove any loops. This change in routing
alleviated the twisting stress and returned the overall transmission
of all affected fibers back to normal. Fiber transmission, when not
being affected by this stress-induced form of FRD, has been shown to
be very stable \citep{ada11a}. However, over the course of the past
year, the transmission of these affected fibers has been
declining. Attempts at relieving the stress with manipulation and
re-routing of the fiber bundle have failed, and the low transmission of the
affected fibers is now permanent. We note that the fiber routing for
the Mitchell Spectrograph, even after the elimination of the fiber
coil, has numerous places where the bend radius is at or tighter than
the 10~cm limit set for the VIRUS instrument. In \S \ref{discuss} we
give further discussion of the implication of this transmission loss
for VIRUS and other IFU fed spectrographs.

\section{DISCUSSION}\label{discuss}

We give a recap of the tests presented here, then discuss what
implications this work has for both the VIRUS instrument and
other projects which employ fiber optics that experience stress or
motion during science observations or deployment. In summary:

\noindent
1) \emph{We find clear evidence for an increase in FRD due to the
accumulation of wear from repeated motion.} Surprisingly, FRD
due to poor fiber surface polish is effectively washed out by this new
FRD, which likely comes from microfractures formed by repeated
motion. This increase in FRD shows no dependence on wavelength. Although
FRD is found to increase with wear, no loss in overall transmission
was seen in the same fibers.

\noindent
2) Fiber motion leads to low-level ($\le 1.0\%$), transient FRD events in a select
number of fibers. The rate necessary to manifest these FRD events
is typical of a telescope at a slew rate of travel (e.g. 40 to 120
mm/sec). When the fibers were moved at a rate typical of a science
observation (e.g. 0.8 to 1.3 mm/sec), the transient FRD events dropped
to a negligible level. These FRD events tended to cluster into bimodal
or trimodal distributions along the the limits of the telescope track
where the stress on the fiber bundle was at a maximum. 

\noindent
3) Bend radius tests on the 266 $\mu$m core fibers showed no evidence
for increased FRD down to a R~=~10~cm bend radius. From
5~cm~$\ge$~R~$\ge$~3~cm, the fibers experienced FRD for slower input
f-ratios ($\sim$f/8) as evidenced by the increase in light thrown into
the central obscuration, but not the outer halo. By R~=~1.5~cm, FRD
increased for both slow and fast (f/3.65) input f-ratios.

\noindent
4) An explosive form of FRD has been seen in the current fiber
bundle in use for the Mitchell Spectrograph. Transmission in 18
fibers is affected, and stems from localized stress at the input
head. The FRD is causing light to scatter both into the central
obscuration and the outer halo. In the most severe cases, when
transmission loss exceeds 50\%, loss of total internal reflection
within the fiber is suspected.

It is instructive to review the causes of these various forms of FRD
before discussing the implications for the VIRUS instrument. We find
evidence for all three distinct forms of FRD as studied in \citet{hay11}. The first is end-surface scattering, due to
poor fiber polish. This can take the form of visual scratches and
pits, and also sub-surface damage from the polishing process. The
evidence for this form of FRD is best seen in Figure \ref{before} where a
clear bimodal distribution in the FRD characteristics of the lifetime
test fiber bundle is seen. The bimodal nature of the FRD curves has an
exact correlation with where on the input head the fiber is located. A
large region of the fiber head was polished to a lower level, and this
is directly reflected in the increased FRD measured of those fibers.

A second form of FRD were the transient events seen throughout the
lifetime tests. As the fibers were in motion at a rate of travel high
enough to not allow for a smooth dissipation of regions of localized
pressure, microbends develop and lead to low levels of FRD. This is
seen in the increased SOE values during the accelerated sections
of the tests, and most prominent at the limits of the tracks where fiber bundle
stress was likely the highest (see Figures \ref{soe1a} and \ref{soe1b}
and in the Appendix). While these FRD events in the lifetime tests were 
transient, we also find that an accumulation of localized shear and
subsequent microbends can become permanent, as seen in the degradation
of the VP2 fiber bundle in use on the Mitchell Spectrograph.

The third form of FRD seen in these tests is evidenced by comparing
the FRD values from before and after the lifetime tests (see Figure
\ref{frdafter}). Although we can not be
certain, we believe this increase in FRD is due to microfractures that
have formed in the fibers from repeated motion. The FRD caused by
repeated flexure is permanent and washes out the FRD due to poor
surface polish (see Figure \ref{frdafter}a).
Of these various sources of FRD, all but one is directly under the
control of the fiber bundle manufacturer. FRD induced by surface
scattering is remedied by higher quality end polish. The
cross-sectional plots shown in Figure \ref{bend} give an indication of
how well a fiber can be polished. However, this was for a single
fiber, and an accurate, flat polish of the much larger surface area of
typical fiber bundle input head is more difficult to achieve than for
the single fiber shown in Figure \ref{bend}.

Perhaps the source of FRD most critical to the VIRUS instrument, and
astronomical instrumentation in general, is the FRD coming from
localized shear and subsequent microbending. The importance of
understanding this form of FRD becomes clear when one considers its
variable nature. Although FRD due to surface scattering is not ideal,
it is at least stable over time. Yet when FRD is variable,
calibrations become difficult to impossible, and systematic errors with
flat-fielding and extraction of on-chip fiber profiles can dominate
your uncertainties. For the VIRUS instrument, an understanding of
these challenges, as discovered from the VIRUS prototype, the bend
radius tests, and the lifetime fiber tests, has led to careful
consideration in the design of the fiber routing for VIRUS. This
includes a bend-radius limit and the design of the three strain
reliefs used to manage the fiber bundles once deployed on the HET.
The current calibration plan for the HETDEX project involves
collecting flat-field and arc lamp frames while slewing the telescope
to target the next field. The levels of transient FRD seen in the
lifetime fiber tests were quite low, typically $\le 0.5\%$ with a
maximum rarely greater than 1.0\%. Whether
this affects the calibrations at a detectable level will need to be
explored further, once VIRUS is on-sky.

A final lesson gleaned from the degradation seen in VP2 is both in the
initial design and packing fraction of the fiber bundle, and how the
fiber bundles are deployed. In VP2, the fiber packing
fraction inside the fiber conduit is tight. Allowing enough space, or
providing a low-friction buffer, such as a Teflon tubing, so
that torsion placed on the fiber conduit does not pass directly to the
fibers, can mitigate the transfer of shear into the fibers. Yet perhaps
more critical is how the fiber bundles are deployed. The coiling of
VP2 was certainly the source of the initial problems seen with fiber
transmission loss. Once the twists of the coiling were released, the
transmission returned to normal for $\sim 2$ years. It is unclear what
has led to the permanent decline in these fibers (Figure
\ref{vp2}), but as the transmission loss was previously shown to come
from localized shear, it's likely that the fibers have worked themselves
into a configuration that can not be undone by fiber bundle
manipulation and shaking. It is also not clear
if the effect seen in VP2 is really a risk for VIRUS, as the bundles
will not experience the relatively sharp bends (R~$\le 10$~cm) found
in the Mitchell Spectrograph, and the VIRUS IFUs will use a Teflon
inner lining. One possible solution for other IFU-based instruments would be
to include a row of dummy fibers along the outer edges of the
IFU. As the twists that led to explosive FRD in VP2 were placed $\sim
10$ meters from the input head, the dummy fibers would need to run the
length of the fiber bundle. For the VIRUS instrument, this was
prohibited by space constraints at the IFU head, but could be
implemented on other IFU configurations for future instrumentation.

\clearpage

\bibliography{fibers12}

\begin{thebibliography}{10}

\bibitem{ang77}
{Angel}, J.~R.~P., {Adams}, M.~T., {Boroson}, T.~A., and {Moore}, R.~L., ``{A
  very large optical telescope array linked with fused silica fibers},'' {\em
  The Astrophysical Journal}~{\bf 218},  776--782 (Dec. 1977).

\bibitem{ram88}
{Ramsey}, L.~W., ``{Focal ratio degradation in optical fibers of astronomical
  interest},'' in [{\em Fiber Optics in
  Astronomy}{\nolinebreak\hspace{0.1em}]},  {Barden}, S.~C., ed., {\em
  Astronomical Society of the Pacific Conference Series} {\bf 3},  26--39
  (1988).

\bibitem{sch03}
{Schmoll}, J., {Roth}, M.~M., and {Laux}, U., ``{Statistical Test of Optical
  Fibers for Use in PMAS, the Potsdam Multi-Aperture Spectrophotometer},'' {\em
  Astronomical Soc. of the Pacific}~{\bf 115},  854--868 (July 2003).

\bibitem{cra08}
{Crause}, L., {Bershady}, M., and {Buckley}, D., ``{Investigation of focal
  ratio degradation in optical fibres for astronomical instrumentation},'' in
  [{\em Society of Photo-Optical Instrumentation Engineers (SPIE) Conference
  Series}{\nolinebreak\hspace{0.1em}]},  {\em Society of Photo-Optical
  Instrumentation Engineers (SPIE) Conference Series} {\bf 7014} (Aug. 2008).

\bibitem{pop10a}
{Poppett}, C.~L. and {Allington-Smith}, J.~R., ``{The dependence of the
  properties of optical fibres on length},'' {\em Mon. Not. R. astr. Soc.}~{\bf
  404},  1349--1354 (May 2010).

\bibitem{cra88}
{Craig}, W.~W., {Hailey}, C.~J., and {Brodie}, J.~P., ``{Measurement of fibers
  to be used in fiber fed spectroscopy},'' in [{\em Fiber Optics in
  Astronomy}{\nolinebreak\hspace{0.1em}]},  {Barden}, S.~C., ed., {\em
  Astronomical Society of the Pacific Conference Series} {\bf 3},  41--51
  (1988).

\bibitem{cla89}
{Clayton}, C.~A., ``{The implications of image scrambling and focal ratio
  degradation in fibre optics on the design of astronomical instrumentation},''
  {\em Astronomical Soc. of the Pacific}~{\bf 213},  502--515 (Apr. 1989).

\bibitem{avi98}
{Avila}, G., ``{Results on Fiber Characterization at ESO},'' in [{\em Fiber
  Optics in Astronomy III}{\nolinebreak\hspace{0.1em}]},  {Arribas}, S.,
  {Mediavilla}, E., and {Watson}, F., eds., {\em Astronomical Society of the
  Pacific Conference Series} {\bf 152},  44--+ (1998).

\bibitem{bry10}
{Bryant}, J.~J., {O'Byrne}, J.~W., {Bland-Hawthorn}, J., and {Leon-Saval},
  S.~G., ``{Hexabundles: first results},'' in [{\em Society of Photo-Optical
  Instrumentation Engineers (SPIE) Conference
  Series}{\nolinebreak\hspace{0.1em}]},  {\em Society of Photo-Optical
  Instrumentation Engineers (SPIE) Conference Series} {\bf 7735} (July 2010).

\bibitem{hay11}
{Haynes}, D.~M., {Withford}, M.~J., {Dawes}, J.~M., {Lawrence}, J.~S., and
  {Haynes}, R., ``{Relative contributions of scattering, diffraction and modal
  diffusion to focal ratio degradation in optical fibres},'' {\em Mon. Not. R.
  astr. Soc.}~{\bf 414},  253--263 (June 2011).

\bibitem{bry11}
{Bryant}, J.~J., {O'Byrne}, J.~W., {Bland-Hawthorn}, J., and {Leon-Saval},
  S.~G., ``{Characterization of hexabundles: initial results},'' {\em Mon. Not.
  R. astr. Soc.}~{\bf 415},  2173--2181 (Aug. 2011).

\bibitem{ber04}
{Bershady}, M.~A., {Andersen}, D.~R., {Harker}, J., {Ramsey}, L.~W., and
  {Verheijen}, M.~A.~W., ``{SparsePak: A Formatted Fiber Field Unit for the
  WIYN Telescope Bench Spectrograph. I. Design, Construction, and
  Calibration},'' {\em Astronomical Soc. of the Pacific}~{\bf 116},  565--590
  (June 2004).

\bibitem{kel04}
{Kelz}, A., {Verheijen}, M., {Roth}, M.~M., {Laux}, U., and {Bauer}, S.-M.,
  ``{Development of the wide-field IFU PPak},'' in [{\em Ground-based
  Instrumentation for Astronomy. Edited by Alan F. M. Moorwood and Iye
  Masanori. Proceedings of the SPIE, Volume 5492, pp. 719-730
  (2004).}{\nolinebreak\hspace{0.1em}]},  {Moorwood}, A.~F.~M. and {Iye}, M.,
  eds., {\em Presented at the Society of Photo-Optical Instrumentation
  Engineers (SPIE) Conference} {\bf 5492},  719--730 (Sept. 2004).

\bibitem{smi04}
{Smith}, G.~A., {Saunders}, W., {Bridges}, T., {Churilov}, V., {Lankshear}, A.,
  {Dawson}, J., {Correll}, D., {Waller}, L., {Haynes}, R., and {Frost}, G.,
  ``{AAOmega: a multipurpose fiber-fed spectrograph for the AAT},'' in [{\em
  Society of Photo-Optical Instrumentation Engineers (SPIE) Conference
  Series}{\nolinebreak\hspace{0.1em}]},  {A.~F.~M.~Moorwood \& M.~Iye}, ed.,
  {\em Society of Photo-Optical Instrumentation Engineers (SPIE) Conference
  Series} {\bf 5492},  410--420 (Sept. 2004).

\bibitem{rot05}
{Roth}, M.~M., {Kelz}, A., {Fechner}, T., {Hahn}, T., {Bauer}, S.-M., {Becker},
  T., {B{\"o}hm}, P., {Christensen}, L., {Dionies}, F., {Paschke}, J., {Popow},
  E., {Wolter}, D., {Schmoll}, J., {Laux}, U., and {Altmann}, W., ``{PMAS: The
  Potsdam Multi-Aperture Spectrophotometer. I. Design, Manufacture, and
  Performance},'' {\em Publications of the Astronomical Society of the
  Pacific}~{\bf 117},  620--642 (June 2005).

\bibitem{kel06b}
{Kelz}, A., {Verheijen}, M.~A.~W., {Roth}, M.~M., {Bauer}, S.~M., {Becker}, T.,
  {Paschke}, J., {Popow}, E., {S{\'a}nchez}, S.~F., and {Laux}, U., ``{PMAS:
  The Potsdam Multi-Aperture Spectrophotometer. II. The Wide Integral Field
  Unit PPak},'' {\em Publications of the Astronomical Society of the
  Pacific}~{\bf 118},  129--145 (Jan. 2006).

\bibitem{tut08}
{Tuttle}, S.~E., {Schiminovich}, D., {Milliard}, B., {Grange}, R., {Martin},
  D.~C., {Rahman}, S., {Deharveng}, J.-M., {McLean}, R., {Tajiri}, G., and
  {Matuszewski}, M., ``{The FIREBall fiber-fed UV spectrograph},'' in [{\em
  Society of Photo-Optical Instrumentation Engineers (SPIE) Conference
  Series}{\nolinebreak\hspace{0.1em}]},  {\em Society of Photo-Optical
  Instrumentation Engineers (SPIE) Conference Series} {\bf 7014} (Aug. 2008).

\bibitem{wil10}
{Wilson}, J.~C., {Hearty}, F., {Skrutskie}, M.~F., {Majewski}, S., {Schiavon},
  R., {Eisenstein}, D., {Gunn}, J., {Blank}, B., {Henderson}, C., {Smee}, S.,
  {Barkhouser}, R., {Harding}, A., {Fitzgerald}, G., {Stolberg}, T., {Arns},
  J., {Nelson}, M., {Brunner}, S., {Burton}, A., {Walker}, E., {Lam}, C.,
  {Maseman}, P., {Barr}, J., {Leger}, F., {Carey}, L., {MacDonald}, N.,
  {Horne}, T., {Young}, E., {Rieke}, G., {Rieke}, M., {O'Brien}, T., {Hope},
  S., {Krakula}, J., {Crane}, J., {Zhao}, B., {Carr}, M., {Harrison}, C.,
  {Stoll}, R., {Vernieri}, M.~A., {Holtzman}, J., {Shetrone}, M.,
  {Allende-Prieto}, C., {Johnson}, J., {Frinchaboy}, P., {Zasowski}, G.,
  {Bizyaev}, D., {Gillespie}, B., and {Weinberg}, D., ``{The Apache Point
  Observatory Galactic Evolution Experiment (APOGEE) high-resolution
  near-infrared multi-object fiber spectrograph},'' in [{\em Society of
  Photo-Optical Instrumentation Engineers (SPIE) Conference
  Series}{\nolinebreak\hspace{0.1em}]},  {\em Society of Photo-Optical
  Instrumentation Engineers (SPIE) Conference Series} {\bf 7735} (July 2010).

\bibitem{nav10}
{Navarro}, R., {Chemla}, F., {Bonifacio}, P., {Flores}, H., {Guinouard}, I.,
  {Huet}, J.-M., {Puech}, M., {Royer}, F., {Pragt}, J.~H., {Wulterkens}, G.,
  {Sawyer}, E.~C., {Caldwell}, M.~E., {Tosh}, I.~A.~J., {Whalley}, M.~S.,
  {Woodhouse}, G.~F.~W., {Span{\`o}}, P., {di Marcantonio}, P., {Andersen},
  M.~I., {Dalton}, G.~B., {Kaper}, L., and {Hammer}, F., ``{Project overview of
  OPTIMOS-EVE: the fibre-fed multi-object spectrograph for the E-ELT},'' in
  [{\em Society of Photo-Optical Instrumentation Engineers (SPIE) Conference
  Series}{\nolinebreak\hspace{0.1em}]},  {\em Society of Photo-Optical
  Instrumentation Engineers (SPIE) Conference Series} {\bf 7735} (July 2010).

\bibitem{sau10}
{Saunders}, W., {Colless}, M., {Saunders}, I., {Hopkins}, A., {Goodwin}, M.,
  {Heijmans}, J., {Brzeski}, J., and {Farrell}, T., ``{MANIFEST: a
  many-instrument fiber-positioning system for GMT},'' in [{\em Society of
  Photo-Optical Instrumentation Engineers (SPIE) Conference
  Series}{\nolinebreak\hspace{0.1em}]},  {\em Society of Photo-Optical
  Instrumentation Engineers (SPIE) Conference Series} {\bf 7735} (July 2010).

\bibitem{sch98}
{Schmoll}, J., {Popow}, E., and {Roth}, M.~M., ``{FRD Optimization for PMAS},''
  in [{\em Fiber Optics in Astronomy III}{\nolinebreak\hspace{0.1em}]},
  {Arribas}, S., {Mediavilla}, E., and {Watson}, F., eds., {\em Astronomical
  Society of the Pacific Conference Series} {\bf 152},  64--+ (1998).

\bibitem{car94}
{Carrasco}, E. and {Parry}, I.~R., ``{A method for determining the focal ratio
  degradation of optical fibres for astronomy},'' {\em Mon. Not. R. astr.
  Soc.}~{\bf 271},  1--+ (Nov. 1994).

\bibitem{mur08}
{Murphy}, J.~D., {MacQueen}, P.~J., {Hill}, G.~J., {Grupp}, F., {Kelz}, A.,
  {Palunas}, P., {Roth}, M., and {Fry}, A., ``{Focal ratio degradation and
  transmission in VIRUS-P optical fibers},'' in [{\em Society of Photo-Optical
  Instrumentation Engineers (SPIE) Conference
  Series}{\nolinebreak\hspace{0.1em}]},  {\em Society of Photo-Optical
  Instrumentation Engineers (SPIE) Conference Series} {\bf 7018} (July 2008).

\bibitem{hay08}
{Haynes}, D.~M., {Withford}, M.~J., {Dawes}, J.~M., {Haynes}, R., and
  {Bland-Hawthorn}, J., ``{Focal ratio degradation: a new perspective},'' in
  [{\em Society of Photo-Optical Instrumentation Engineers (SPIE) Conference
  Series}{\nolinebreak\hspace{0.1em}]},  {\em Society of Photo-Optical
  Instrumentation Engineers (SPIE) Conference Series} {\bf 7018} (July 2008).

\bibitem{bru10}
{Brunner}, S., {Burton}, A., {Crane}, J., {Zhao}, B., {Hearty}, F.~R.,
  {Wilson}, J.~C., {Carey}, L., {Leger}, F., {Skrutskie}, M., {Schiavon}, R.,
  and {Majewski}, S.~R., ``{APOGEE fiber development and FRD testing},'' in
  [{\em Society of Photo-Optical Instrumentation Engineers (SPIE) Conference
  Series}{\nolinebreak\hspace{0.1em}]},  {\em Society of Photo-Optical
  Instrumentation Engineers (SPIE) Conference Series} {\bf 7735} (July 2010).

\bibitem{pop10b}
{Poppett}, C. and {Allington-Smith}, J., ``{A new method to quantitatively
  compare focal ratio degradation due to different end termination
  techniques},'' in [{\em Society of Photo-Optical Instrumentation Engineers
  (SPIE) Conference Series}{\nolinebreak\hspace{0.1em}]},  {\em Society of
  Photo-Optical Instrumentation Engineers (SPIE) Conference Series} {\bf 7735}
  (July 2010).

\bibitem{hil10}
{Hill}, G.~J., {Lee}, H., {Vattiat}, B.~L., {Adams}, J.~J., {Marshall}, J.~L.,
  {Drory}, N., {Depoy}, D.~L., {Blanc}, G., {Bender}, R., {Booth}, J.~A.,
  {Chonis}, T., {Cornell}, M.~E., {Gebhardt}, K., {Good}, J., {Grupp}, F.,
  {Haynes}, R., {Kelz}, A., {MacQueen}, P.~J., {Mollison}, N., {Murphy}, J.~D.,
  {Rafal}, M.~D., {Rambold}, W.~N., {Roth}, M.~M., {Savage}, R., and {Smith},
  M.~P., ``{VIRUS: a massively replicated 33k fiber integral field spectrograph
  for the upgraded Hobby-Eberly Telescope},'' in [{\em Society of Photo-Optical
  Instrumentation Engineers (SPIE) Conference
  Series}{\nolinebreak\hspace{0.1em}]},  {\em Society of Photo-Optical
  Instrumentation Engineers (SPIE) Conference Series} {\bf 7735} (July 2010).

\bibitem{hil08a}
{Hill}, G.~J., {Gebhardt}, K., {Komatsu}, E., {Drory}, N., {MacQueen}, P.~J.,
  {Adams}, J.~J., {Blanc}, G.~A., {Koehler}, R., {Rafal}, M., {Roth}, M.~M.,
  {Kelz}, A., {Gronwall}, C., {Ciardullo}, R., and {Schneider}, D.~P., ``{The
  Hobby-Eberly Telescope Dark Energy Experiment (HETDEX): Description and Early
  Pilot Survey Results},'' in [{\em {Panoramic Views of the
  Universe}}{\nolinebreak\hspace{0.1em}]},  {\em ASP conference series} (2008).

\bibitem{kel06}
{Kelz}, A., {Bauer}, S.~M., {Grupp}, F., {Hill}, G.~J., {Popow}, E., {Palunas},
  P., {Roth}, M.~M., {MacQueen}, P.~J., and {Tripphahn}, U., ``{Prototype
  development of the integral-field unit for VIRUS},'' in [{\em Optomechanical
  Technologies for Astronomy. Edited by Atad-Ettedgui, Eli; Antebi, Joseph;
  Lemke, Dietrich. Proceedings of the SPIE, Volume 6273, pp. 62733W
  (2006).}{\nolinebreak\hspace{0.1em}]},  {\em Presented at the Society of
  Photo-Optical Instrumentation Engineers (SPIE) Conference} {\bf 6273} (July
  2006).

\bibitem{hil08b}
{MacQueen}, P.~J., {Hill}, G.~J., {Smith}, M.~P., {Tufts}, J.~R., {Barnes},
  S.~I., {Roth}, M.~M., {Kelz}, A., {Adams}, J.~J., {Blanc}, G., {Murphy},
  J.~D., {Altmann}, W., {Wesley}, G.~L., {Segura}, P.~R., {Good}, J.~M.,
  {Goertz}, J.~A., {Edmonston}, R.~D., and {Wilkinson}, C.~P., ``{Design,
  construction, and performance of VIRUS-P: the prototype of a highly
  replicated integral field spectrograph for the HET},'' in [{\em Astronomical
  Telescopes and Instrumentation}{\nolinebreak\hspace{0.1em}]},  {\em Proc.
  SPIE} {\bf 7014-257} (2008).

\bibitem{sou10}
{Soukup}, I.~M., {Beno}, J.~H., {Hayes}, R.~J., {Heisler}, J.~T., {Mock},
  J.~R., {Mollison}, N.~T., {Good}, J.~M., {Hill}, G.~J., {Vattiat}, B.~L.,
  {Murphy}, J.~D., {Anderson}, S.~C., {Bauer}, S.~M., {Kelz}, A., {Roth},
  M.~M., and {Fahrenthold}, E.~P., ``{Design of the fiber optic support system
  and fiber bundle accelerated life test for VIRUS},'' in [{\em Society of
  Photo-Optical Instrumentation Engineers (SPIE) Conference
  Series}{\nolinebreak\hspace{0.1em}]},  {\em Society of Photo-Optical
  Instrumentation Engineers (SPIE) Conference Series} {\bf 7735} (July 2010).

\bibitem{lee10}
{Lee}, H., {Hill}, G.~J., {Marshall}, J.~L., {Vattiat}, B.~L., and {Depoy},
  D.~L., ``{Visible Integral-field Replicable Unit Spectrograph (VIRUS) optical
  tolerance},'' in [{\em Society of Photo-Optical Instrumentation Engineers
  (SPIE) Conference Series}{\nolinebreak\hspace{0.1em}]},  {\em Society of
  Photo-Optical Instrumentation Engineers (SPIE) Conference Series} {\bf 7735}
  (July 2010).

\bibitem{pop07}
{Poppett}, C.~L. and {Allington-Smith}, J.~R., ``{Fibre systems for future
  astronomy: anomalous wavelength-temperature effects},'' {\em Mon. Not. R.
  astr. Soc.}~{\bf 379},  143--150 (July 2007).

\bibitem{ada11a}
{Adams}, J.~J., {Blanc}, G.~A., {Hill}, G.~J., {Gebhardt}, K., {Drory}, N.,
  {Hao}, L., {Bender}, R., {Byun}, J., {Ciardullo}, R., {Cornell}, M.~E.,
  {Finkelstein}, S.~L., {Fry}, A., {Gawiser}, E., {Gronwall}, C., {Hopp}, U.,
  {Jeong}, D., {Kelz}, A., {Kelzenberg}, R., {Komatsu}, E., {MacQueen}, P.~J.,
  {Murphy}, J., {Odoms}, P.~S., {Roth}, M., {Schneider}, D.~P., {Tufts}, J.~R.,
  and {Wilkinson}, C.~P., ``{The HETDEX Pilot Survey. I. Survey Design,
  Performance, and Catalog of Emission-line Galaxies},'' {\em The Astrophysical
  Journal, Supplement}~{\bf 192},  5 (Jan. 2011).

\end{thebibliography}
\bibliographystyle{apj}

\section{Appendix}

\begin{figure}
\begin{center}
\begin{tabular}{c}
\includegraphics[height=92mm,width=160mm]{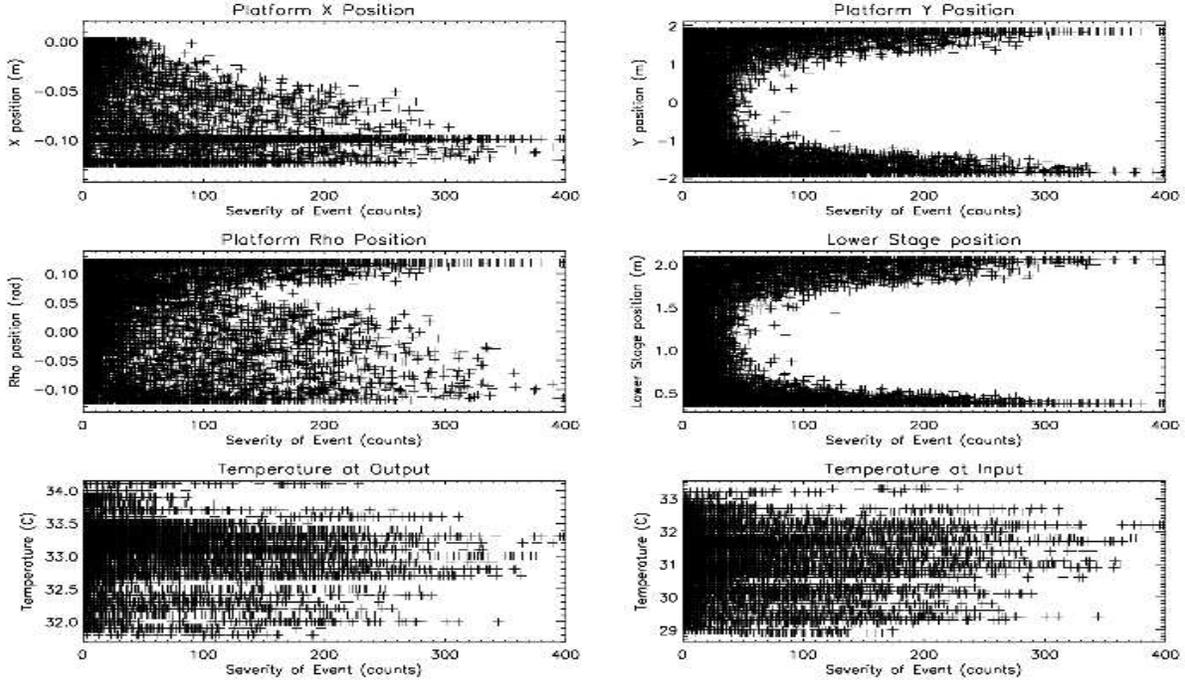}
\end{tabular}
\end{center}
\caption[The SOE values for the Dec 38 track in slew mode (rate
  = 140 mm/sec).]{The SOE values for the Dec 38 track in slew mode (rate
  = 140 mm/sec). Symbols and axes are the same as Figure
  \ref{soe1a}. There are 9,896 divided frames plotted. Note that for
  this track, the height of the lower bench and the Y position are 
  synchronized, thus the clear trend of FRD events occurring at the
  extreme positions of both the Y motion and lower stage. Further
  testing showed that most of these events were driven by motion of
  the lower stage. Here we again see a trend with temperature,
  although the trend is a weak one.
  \label{soe2a}}
\end{figure}

\begin{figure}
\begin{center}
\begin{tabular}{c}
\includegraphics[height=92mm,width=160mm]{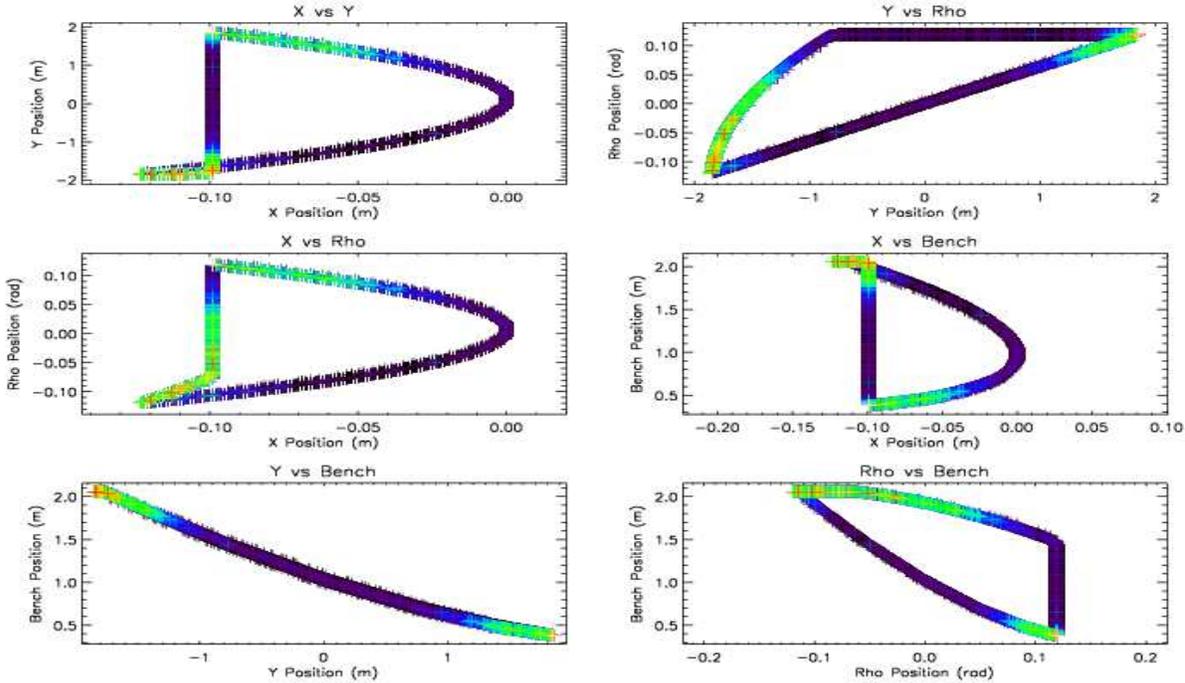}
\end{tabular}
\end{center}
\caption[The severity-of-event values plotted against the
  4 motion parameters, X, Y, rho, and the lower bench height.]
  {The SOE values from Figure \ref{soe2a} plotted against the
  4 motion parameters, X, Y, rho, and the lower bench height for the
  range of SOE values plotted in Figure \ref{soe2a}. Red plots high
  SOE values while black plots SOE values of zero. See Figure
  \ref{soe1b} for a more complete description. Here we see that the Y
  and lower stage motion is driving the FRD events, with the X and rho
  motion showing less influence.
  \label{soe2b}} 
\end{figure}

\begin{figure}
\begin{center}
\begin{tabular}{c}
\includegraphics[height=95mm,width=160mm]{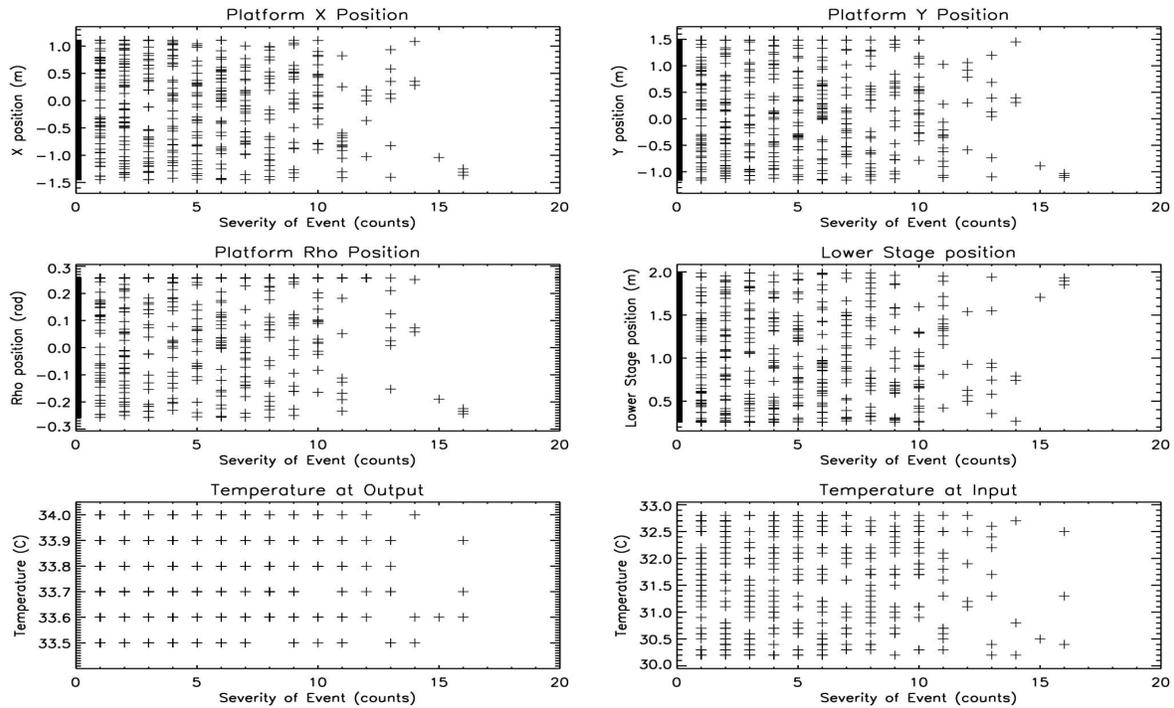}
\end{tabular}
\end{center}
\caption[The SOE values (1,321 divided frames) for the Dec 38 track
  in observation mode (rates of 0.77 to 1.3 mm/sec).]{The SOE values
  (1,321 divided frames) for the Dec 38 track 
  in observation mode (rates of 0.77 to 1.3 mm/sec). Note the much
  smaller range of SOE values plotted here.
  \label{soe4a}}
\end{figure}

\begin{figure}
\begin{center}
\begin{tabular}{c}
\includegraphics[height=95mm,width=160mm]{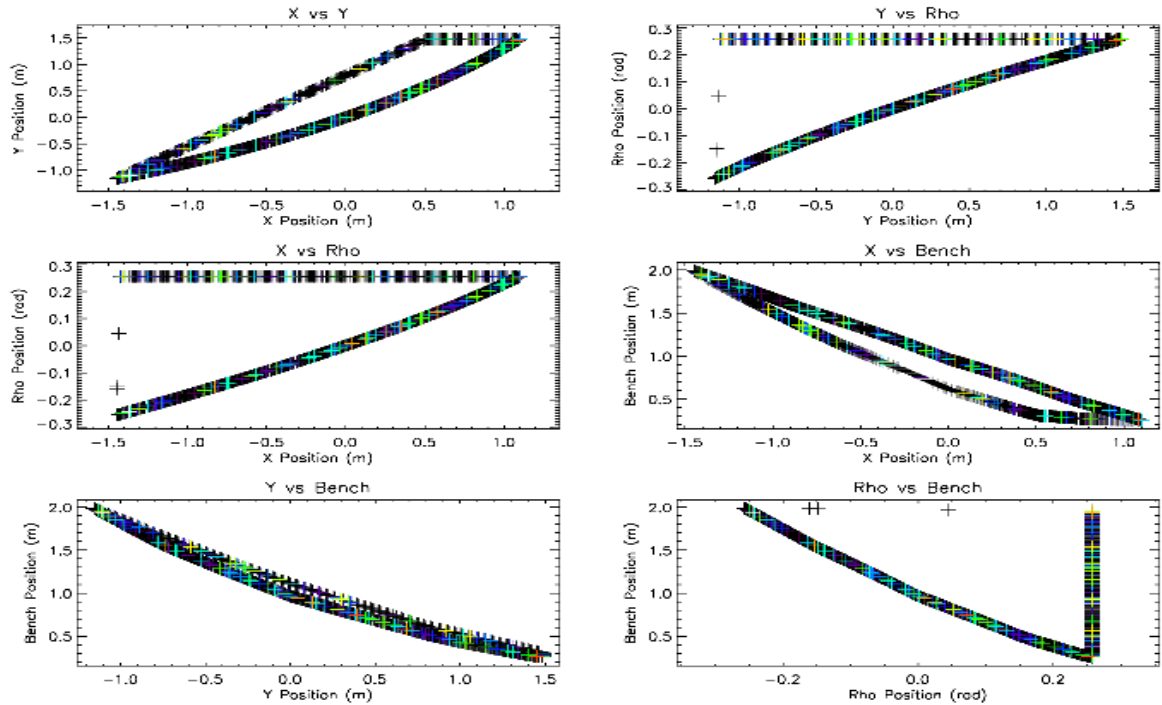}
\end{tabular}
\end{center}
\caption[The SOE values plotted against the 4 parameters of
  motion.]{The SOE values plotted against the 4 parameters of motion 
  as described in Figure \ref{soe1b}. The range of SOE values is
  significantly smaller (by a factor of $\sim 20 \times$) than other
  tracks run in slew mode. Also note that the high SOE values are
  scattered evenly along the entire track. This pattern is in marked
  contrast to the tests in slew mode where FRD events are clearly
  clustered. The range of this track was slightly altered from the
  other Dec 38 track (Figures \ref{soe2a} and \ref{soe2b}) in order to
  test the observation rate and a slew rate rewind (note the
  undersampling in rho during the rapid rewind period).
  \label{soe4b}}
\end{figure}

\begin{figure}
\begin{center}
\begin{tabular}{c}
\includegraphics[height=95mm,width=160mm]{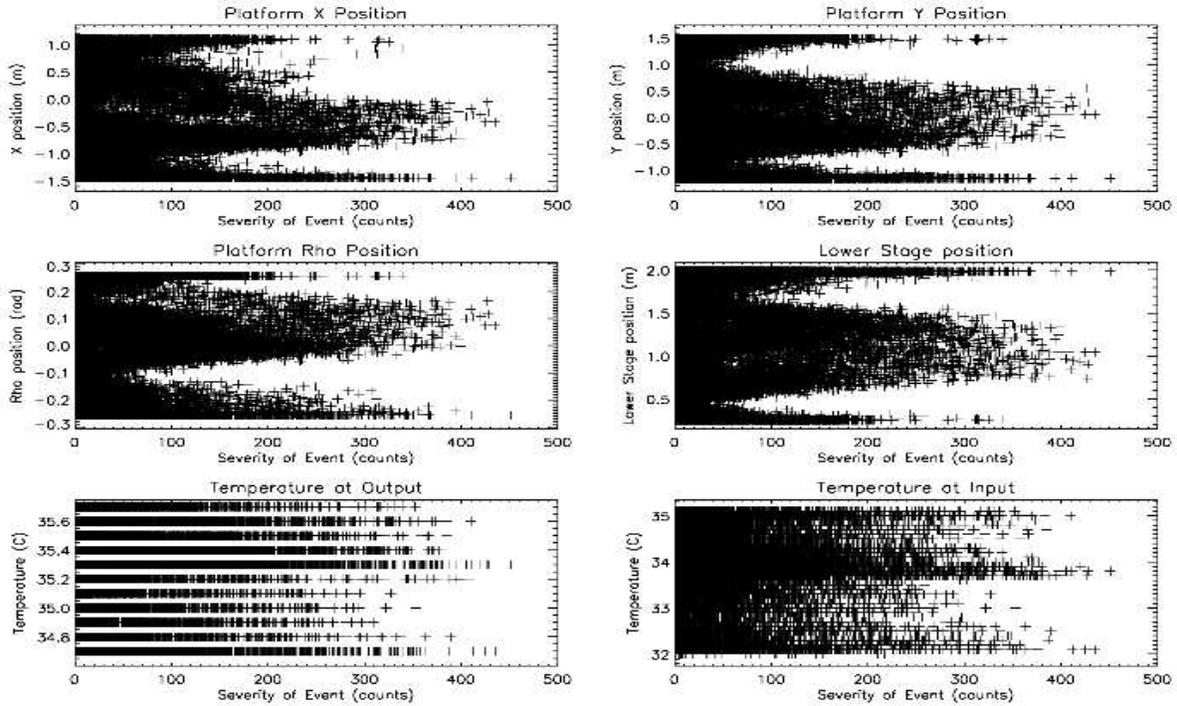}
\end{tabular}
\end{center}
\caption[SOE values for the Dec 60 track in slew mode.]{SOE values for
  the Dec 60 track in slew mode. These  
  24,545 frames were all taken at a rate of 105 mm/sec. Note that the
  SOE range plotted in this figure has increased from 400 to 500. Here
  we see a trimodal distribution in the locations of the most severe
  events. Also note that there is no clear trend with temperature as
  seen in Figures \ref{soe1a} and \ref{soe2a}.
  \label{soe3a}}
\end{figure}

\begin{figure}
\begin{center}
\begin{tabular}{c}
\includegraphics[height=95mm,width=160mm]{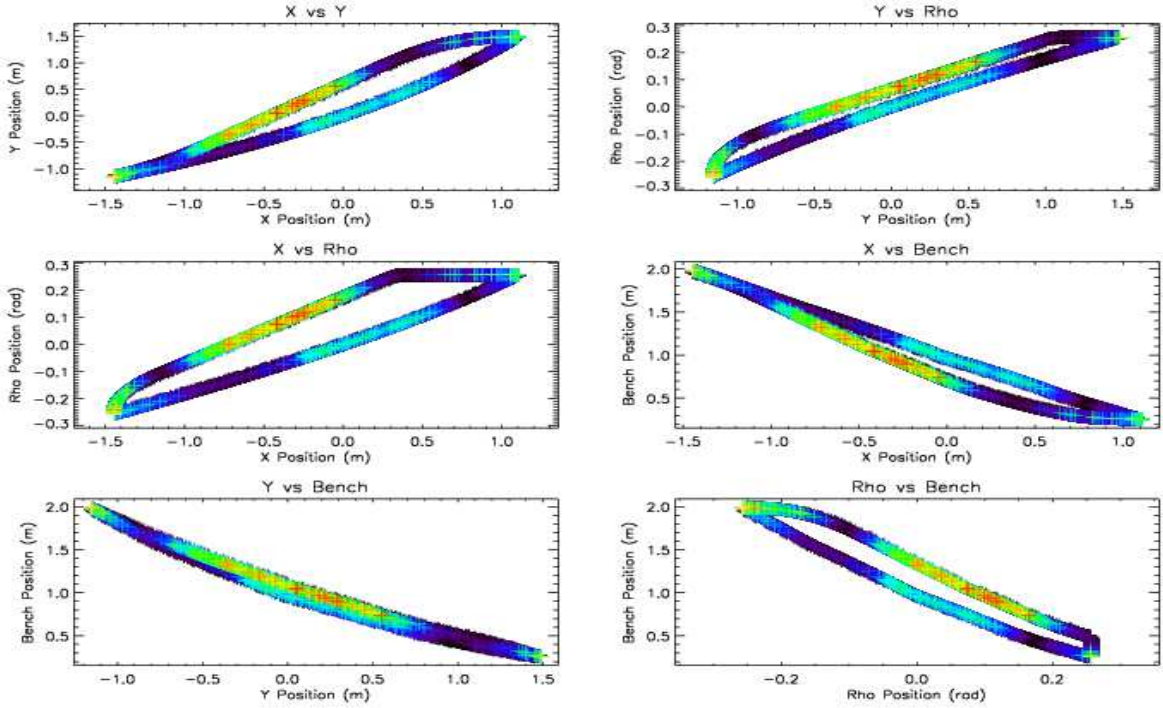}
\end{tabular}
\end{center}
\caption[The SOE values plotted against the 4 parameters of
  motion.]{The SOE values plotted against the 4 parameters of motion 
  as described in Figure \ref{soe1b}. Unlike the other two tracks (Dec
  38 and Dec 65) run in slew mode, we see a trimodal distribution in
  the locations of the most extreme FRD events. These manifest at both
  the limits of X and Y, and also the central position of all 4
  parameters.
  \label{soe3b}}
\end{figure}

\end{document}